\documentclass[pre,10pt,twocolumn,aps,longbibliography,amsmath,amssymb,superscriptaddress]{revtex4-2}
\usepackage{graphicx}
\usepackage{mathtools}

\PassOptionsToPackage{hyphens}{url}\usepackage{hyperref}

\hypersetup{
  colorlinks=true,
  linkcolor=blue,
  urlcolor=blue,
  citecolor=blue
}

\def\rmd {\mathrm{d}} 

\newcommand{\N} {\mathbb{N}} 
 
\newcommand{\R} {\mathbb{R}}

\newcommand{\lrbrace}[1] {\left\{ #1 \right\}}
\DeclarePairedDelimiter{\lr}{(}{)}
\DeclarePairedDelimiter{\lrvert}{\lvert}{\rvert}

\begin{document}

\title{Long-Range Interacting Particles on a Helix: A Statistical and Correlation Analysis of Equilibrium Configurations}

\author{J. M. Dörre}\email{mathisdoerrehh@gmail.com}
  \affiliation{Zentrum für Optische Quantentechnologien, Fachbereich Physik, Universität Hamburg, Luruper Chaussee 149, 22761 Hamburg Germany}
\author{F. K. Diakonos}
  \affiliation{Department of Physics, National and Kapodistrian University of Athens, GR-15784 Athens Greece}
\author{P. Schmelcher}\email{peter.schmelcher@uni-hamburg.de}
  \affiliation{Zentrum für Optische Quantentechnologien, Fachbereich Physik, Universität Hamburg, Luruper Chaussee 149, 22761 Hamburg Germany}
  \affiliation{The Hamburg Centre for Ultrafast Imaging, Universität Hamburg, Luruper Chaussee 149, 22761 Hamburg Germany}

\date{\today}

\begin{abstract}
We provide a statistical and correlational analysis of the spatial and energetic properties of equilibrium configurations
of a few-body system of two to eight equally charged classical particles that are confined 
on a one-dimensional helical manifold. The two-body system has been demonstrated to yield an oscillatory
effective potential, thus providing stable equilibrium configurations despite the repulsive
Coulomb interactions. As the system size grows, the number of equilibria increases, approximately
following a power-law.
This can be attributed to the increasing complexity in the highly non-linear oscillatory
behavior of the potential energy surface.
This property is reflected in a crossover from a spatially regular distribution of equilibria
for the two-body system to a heightened degree of disorder upon the addition of particles.
However, in accordance with the repulsion within a helical winding, the
observed interparticle distances in equilibrium configurations cluster around
values of odd multiples of half a helical winding, thus maintaining an underlying regularity.
Furthermore, an energetic hierarchy exists based on the spatial location of the local equilibria,
which is subject to increasing fluctuations as the system size grows.
\end{abstract}

\maketitle

\section{Introduction}\label{chap:1}
One of the structures frequently observed in nature is the helix. The perhaps most prominent example is the deoxyribonucleic acid (DNA) 
double-helix \cite{watsonMolecularStructureNucleic1953}, whose structure is pivotal in enabling the compact packing of the extensive polymer 
while preserving the accessibility of the protein-coding information \cite{traversDNAStructureFunction2015}. Another noteworthy prevalence of a helix 
is the $\alpha$-helix, which is nearly ubiquitous as the secondary structure of the polypeptide chains in proteins 
\cite{paulingStructureProteinsTwo1951,paulingConfigurationsPolypeptideChains1951,chothiaStructureProteinsPacking1977,cohenAlphahelicalCoiledCoils1990}. 
Their helical structures render both $\alpha$-helices and DNA, potentially suitable as spin filters 
\cite{gutierrezSpinselectiveTransportHelical2012,guoSpinSelectiveTransportElectrons2012,guoSpindependentElectronTransport2014} or field-effect 
transistors \cite{yooElectricalConductionPolydAPolydT2001,malyshevDNADoubleHelices2007} in the domain of molecular electronics.

Driven by the functionalities of helices in nature, fabrication methods and properties of helical nanostructures have been studied in 
detail \cite{renReviewHelicalNanostructures2014}. In addition to fully guided manufacturing methods \cite{huangHelicesMicroworldMaterials2015}, 
the research has yielded self-assembled helical structures, for instance, due to crystal growth 
\cite{huangHelicesMicroworldMaterials2015,lauCoiledCarbonNanotubes2006,shaikjeeSynthesisPropertiesUses2012,renReviewHelicalNanostructures2014,gibbsNanohelicesShadowGrowth2014}, 
template-assisted approaches \cite{huangHelicesMicroworldMaterials2015,liuHelicalNanostructuresBased2014}, or long-range ordering on cylindrical surfaces 
\cite{kohlstedtSpontaneousChiralityLongrange2007,vernizziElectrostaticOriginChiral2009,srebnikSelfassemblyChargedParticles2011}. Many of these nanohelices 
possess remarkable mechanical, electrical, and optical properties 
\cite{gaoSuperelasticityNanofractureMechanics2006,lauCoiledCarbonNanotubes2006,tascoThreedimensionalNanohelicesChiral2016,kostersCoreShellPlasmonic2017}. 
Notable examples include the superelasticity of ZnO nanohelices \cite{gaoSuperelasticityNanofractureMechanics2006} or the unique capacity of coiled carbon 
nanotubes for electronic property modification based on chirality, which may offer potential applications in nanoelectronic 
devices \cite{lauCoiledCarbonNanotubes2006}. An additional, more fundamental approach has been realized in the form of a double-helical trap for cold neutral 
atoms. This trap utilizes an optical nanofiber surrounded by a two-color evanescent field, thereby establishing a platform for fundamental physical research 
\cite{reitzNanofiberbasedDoublehelixDipole2012}. Furthermore, the arrangement of charged dust grains into quasi-crystal helical 
structures (zigzag, helix, interwoven helices) has been observed in a complex plasma in the context of external potentials and of self-confinement 
by a field of smaller background grains in a cylindrically symmetric setup 
\cite{tsytovichHelicalStructuresComplex2005b,tsytovichHelicalStructuresComplex2005a,kamimuraCoulombDoubleHelical2012}.

The above literature and knowledge serves as a motivation to explore particle systems in a helical configuration.
From a conceptual perspective, there is considerable interest in the study of equally charged classical particles since the
combination of long-range interactions and the helical geometry are responsible 
for novel emerging properties. Indeed, it has been shown that, despite the purely repulsive 
Coulomb interactions in 3D, equally charged classical particles confined to a 1D helical manifold
experience an oscillatory force as a function of the interparticle distance 
\cite{kibisElectronelectronInteractionSpiral1992,schmelcherEffectiveLongrangeInteractions2011}. Consequently, such systems have 
stable equilibrium configurations (ECs), that are tunable by the geometric parameters of the helix 
\cite{schmelcherEffectiveLongrangeInteractions2011}. The investigation of dipolar particles with helical confinement shows similar 
phenomena \cite{pedersenFormationClassicalCrystals2014,pedersenQuantumFewbodyBound2016}. Furthermore, the Coulomb-interacting particle 
system has been studied on a closed toroidal helix and under the influence of external fields 
\cite{siemensTunableOrderHelically2020,plettenbergLocalEquilibriaState2017,gloyDrivenToroidalHelix2022}.

For charged particles on a 
helix, the center of mass motion separates from the relative motion due to the homogeneity of the helix 
\cite{zampetakiClassicalScatteringCharged2013}, rendering it an excellent setup to study the dynamics of the system. By 
introducing an inhomogeneity, such as a local modification of the helix radius, the coupling of the center of mass and the relative 
motion of two particles leads to the dissociation of helical bound states in a scattering process 
\cite{zampetakiClassicalScatteringCharged2013}. Moreover, the collision of many-body bound clusters of equally charged particles 
can lead to their decay into a hierarchy of multiple smaller clusters, which subsequently scatter with other clusters, resulting in 
a decay cascade \cite{siemensClassicalScatteringFragmentation2025}. Further investigations of the dynamics of charged particles on a 
closed toroidal helix have shown that the helix radius plays a critical role, influencing the excitation behavior from dispersion to 
self-focusing \cite{zampetakiDynamicsNonlinearExcitations2015} and causing transitions in the ground state configuration and 
vibrational spectrum, including a critical helix radius where the vibrational frequency spectrum collapses to a single 
frequency \cite{zampetakiDegeneracyInversionBand2015}.

In spite of the existing literature on equally charged classical particles in helical confinement, a comprehensive structural and statistical analysis of the 
spatial and energetic properties of the ECs of few-body bound states (not to mention a many-body system) on the infinitely extended straight helix is lacking. 
In the present study, this analysis is performed for a system of up to eight particles, thus allowing a deeper understanding of the structural intricacies inherent 
in the ECs. Due to the highly complex potential energy surface, numerical optimization approaches have to be employed
thereby ensuring to find all ECs. Our findings reveal that the number of ECs approximately follows a power-law depending on the number of particles. 
We observe that the 
helix, as a homogeneous manifold, gives rise to a spatial structure of ECs that lies between complete regularity and disorder, whereby
the degree of disorder increases as the number of particles increases.
A hierarchy of ECs goes hand in 
hand with a corresponding clusters formation.
The latter is consistent with the regularity observed for the simple two-body systems, where ECs occur for interparticle distances of approximately odd 
multiples of half a helical winding \cite{schmelcherEffectiveLongrangeInteractions2011}. Correlations among interparticle distances
within the ECs exhibit a tendency of neighboring distances to be in the same or neighboring clusters. Further, the interparticle
distances are likely to order hierarchically or symmetrically within the ECs.
The energies corresponding to the ECs exhibit a hierarchy based on the interparticle distances. This hierarchy is disturbed by
the addition of particles, leading to increasing energetical fluctuations, attributable to the complex oscillations of
the potential energy surface.

This work is structured as follows. In section \ref{chap:two} the setup of charged particles confined on a helix is described, followed by a 
derivation of the corresponding Hamiltonian. In addition, a concise summary of the properties of the two-body system is given. 
Section \ref{chap:three} delineates the computational approach and discusses the challenges posed by the irregular high-dimensional potential 
landscape. The results of the analysis are presented and discussed in section \ref{chap:four}. This will be done by analyzing the statistics 
and correlations of the spatial properties of the ECs, followed by their energetic properties. Finally, we will examine the
correlations between the spatial and energetic properties. Section \ref{chap:five}, provides a summary and outlook.

\begin{figure}[t]
    \centering
    \includegraphics[width=\columnwidth]{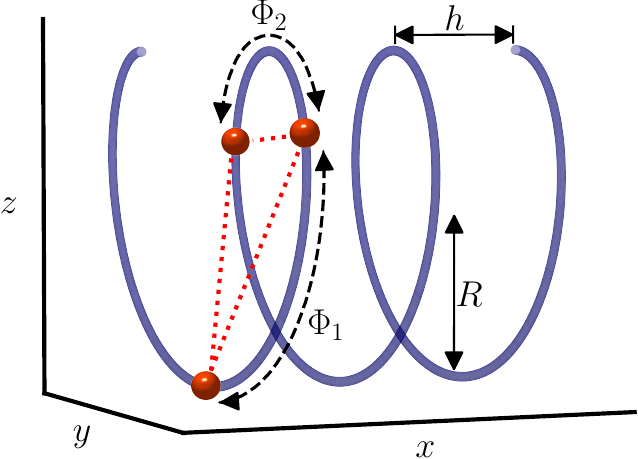}
    \caption{Sketch of three equally charged particles on a helix with radius $R$ and pitch $h$ for $h/R = 0.8$.
        Particles are arranged in an EC, which means that neither particle can move without increasing the
        potential energy. The red dotted lines indicate the repulsive Coulomb interactions.}
    \label{fig:helix}
\end{figure}
\section{Setup, Hamiltonian and two-body problem}\label{chap:two}
We consider a system of $N$ equally charged particles (e.g. ions or electrons) that are confined to a straight, infinite, one-dimensional 
helix. Irrespective of their spatial confinement onto the helix, the particles interact through three-dimensional space via 
Coulomb interactions, forming a mixed-dimensional physical system. This setup is illustrated in Fig. \ref{fig:helix}
for three particles ($N=3$). The Cartesian coordinates of the $i$-th particle on 
the helix are parametrized by
\begin{equation}\label{eq:parametrization}
    \mathbf{r}_i(\phi_i) = \begin{pmatrix}
        x_i(\phi_i) \\
        y_i(\phi_i) \\
        z_i(\phi_i)
    \end{pmatrix} = \begin{pmatrix}
        \frac{h}{2\pi}\phi_i \\
        R\sin(\phi_i) \\
        R\cos(\phi_i)
    \end{pmatrix},
\end{equation}
where $R$ is the radius, $h$ is the pitch of the helix, and $\phi_i$ are its parametric coordinates (azimuthal angles). 
In Cartesian coordinates, the Hamiltonian has the form $\mathcal{H} = \mathcal{T} + \mathcal{V}$, where 
$\mathcal{T} = \sum_{i} \frac{m_i}{2}\dot{\mathbf{r}}_i^2$ is the kinetic energy and 
$\mathcal{V} = \sum_{i<k} g_{ik}/\lrvert*{\mathbf{r}_i - \mathbf{r}_k}$ is the potential energy. We consider 
equal masses $m_i = m$ and equal coupling constants $g_{ik} = g = q^2/(4\pi\varepsilon_0)$. The arc length 
parametrization
\begin{equation}
    s_i: \phi_i \mapsto \int_0^{\phi_i} \lrvert*{\partial\mathbf{r}_i(\phi_i')/\partial\phi_i'}\,\rmd\phi_i'
\end{equation}
yields the proportionality factor $\xi = \sqrt{R^2 + (h/2\pi)^2}$ between the path length and the parametric 
coordinate, i.e., $s_i = \xi\phi_i$. In addition, the definition of $\rho \equiv h/R$ is employed, resulting in the 
following expression of the Hamiltonian:
\begin{align}
    &\mathcal{H} = \sum_{i=1}^{N} \frac{m }{2}\dot{s}_i^2  \notag\\
    &+ \sum_{\substack{i,k = 1 \\ i < k}}^{N} 
    \frac{g/R}{\sqrt{\lr*{\frac{\rho }{2\pi\xi}}^2(s_i - s_k)^2 + 2\lr*{1-\cos\lr*{\frac{1}{\xi}(s_i-s_k)}}}}.
\end{align}
The homogeneity of the helix makes it possible to separate the center of mass from the relative motion.
Therefore, by introducing relative distances $\tilde{s}_i = s_{i+1} - s_i$ between the $i$-th and the $(i+1)$-th 
particle, as well as the center of mass $S = \sum_i ms_i/M$, with the total mass $M = Nm$, the many-body Hamiltonian
can be expressed in internal coordinates:
\begin{widetext}
\begin{equation}\label{eq:hamiltonianinternal}
    \mathcal{H} = \frac{M }{2}\dot{S}^2 + \frac{m}{2}\sum_{i=1}^{N}\lr*{\frac{1}{N}\sum_{k=1 }^{N-1}k\dot{\tilde{s}}_k - \sum_{l=i}^{N-1}\dot{\tilde{s}}_l}^2
    + \sum_{k=1}^{N-1}\sum_{n=k}^{N-1}\frac{g/R}{\sqrt{\lr*{\frac{\rho }{2\pi\xi}}^2 \lr*{\sum_{i=k }^{n}\tilde{s}_i}^2 + 2\lr*{1 - \cos\lr*{\frac{1}{\xi}\sum_{i=k}^{n}\tilde{s}_i}}}}.
\end{equation}
\end{widetext}
Rather than employing path lengths, the more intuitive quantitative measure, $\Phi_i = \tilde{s}_i/(2\pi\xi)$,
which counts the number of helical windings traversed, will be utilized. For the sake of simplicity, we also utilize the 
dimensionless potential energy $V = \mathcal{V}R/g$.

\begin{figure}[t]
    \centering
    \includegraphics[width=\columnwidth]{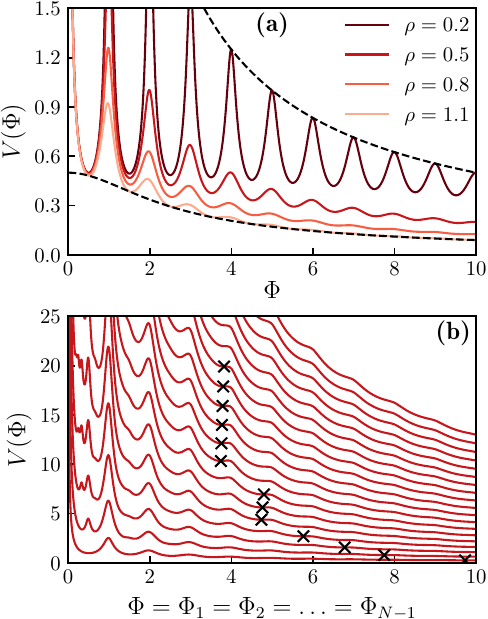}
    \caption{\textbf{(a)} Potential energy $V$ for two charged particles confined on a helix,
    depending on the interparticle distance $\Phi = \Phi_1$. The potential is shown for 
    $\rho=0.2$, $\rho=0.5$, $\rho=0.8$, and $\rho=1.1$. Additionally, the dashed lines show the upper 
    and lower envelope of the potential energy for $\rho=0.2$ and $\rho = 1.1$, respectively. 
    \textbf{(b)} An intersection of the many-body potential landscape $V(\Phi)$, where
    $\Phi = \Phi_1 = \Phi_2 = \ldots = \Phi_{N-1}$, i.e. for equidistant particles, for different $N$ 
    and $\rho = 0.8$. The energetically lowest curve corresponds to $N=2$, the second lowest to $N=3$, etc. The black
    crosses mark the minimum of the outermost potential well of each curve.}
    \label{fig:tbptombp}
\end{figure}

In the case of two repulsively interacting charged particles on the helix ($N=2$), the potential energy term in 
Eq. \eqref{eq:hamiltonianinternal} simplifies to an effective one-body problem. The presence of the cosine 
term in the denominator indicates an oscillatory effective potential and thus the existence of stable ECs for an 
appropriate choice of the geometric parameter $\rho$. 

In the exact limit of $\rho=h=0$, the helix becomes a circle, resulting in ECs on a polygon with equal
interparticle distances. For a helix with finite $h$ the ECs are approximately odd multiples of half a helical winding 
i.e. we have $\Phi = n + \frac{1}{2}$, $n \in \N_0$. In these ECs, the particles are located
approximately on opposite sides of the windings thereby maximizing the Euclidean distance and
consequently minimizing the potential energy. To be more precise, the true minima of the potential energy are located
at values slightly greater than odd multiples of half a helical winding due to the helical winding along its axis. 
Fig. \ref{fig:tbptombp}(a) illustrates the oscillatory behavior of the potential energy for the case of
two particles $N=2$ for different values of the parameter $\rho$, with each minimum corresponding to an EC.
The number of ECs scales approximately as $\lfloor 2\pi/\rho^2 \rfloor$, i.e. a larger ratio $\rho$ leads
to a larger number of stable ECs. Applying the condition $\rmd V/\rmd \Phi = 0$ and $\lvert\sin(2\pi\Phi)\rvert \leq 1$,
yields the upper limit $\Phi \leq 2\pi/\rho^2$, above which the quadratic term in the denominator of Eq.
\eqref{eq:hamiltonianinternal} becomes predominant. Consequently, there are no further minima leading
to stable ECs.
For the ECs near that cutoff value, the deviation from the aforementioned distances of approximately
odd multiples of half a helical winding
is significant, thereby establishing a kind of spatial irregularity for the distances between the ECs.
By increasing $\rho$, this irregularity becomes dominant.

Furthermore, the potential (see Fig.\ref{fig:tbptombp}(a)) exhibits an overall
decrease of its minimum and maximum values belonging to the invididual wells with increasing 
interparticle distance, due to the repulsive three-dimensional interaction. Hence, an energetic hierarchy is formed,
whereby the energies corresponding to the ECs decrease with increasing equilibrium interparticle distances.
The depth of the wells and the stability of the ECs also follow a corresponding hierarchy.
In essence, already the series of ECs of only two equally charged particles on a helix with sufficiently
small $\rho$ show a transition from an approximately regular to a somewhat irregular behavior
with increasing value of the interparticle distance.
Of course, in the limiting case $\rho \to \infty$, the helix becomes a straight line, and therefore 
no ECs occur but any originally nearby placed particles would follow a "Coulomb explosion".

The investigation of the equilibrium structures of helical few-body systems in the range $N=2-8$,
conducted in this work, will further manifest their hierarchical properties between regularity and irregularity
with a focus on their statistical behaviour. To accomplish this we choose for $\rho$ the value $0.8$.
Let us provide a teaser of what awaits us.
Fig. \ref{fig:tbptombp}(b) shows an one-dimensional intersection of the $(N-1)$-dimensional potential
landscape for equidistant particles on the helix, i.e. along its global diagonal 
$\Phi = \Phi_1 = \Phi_2 = \ldots = \Phi_{N-1}$ and for different numbers of particles. Despite the fact that 
this illustration is limited to a low-dimensional intersection of the originally high-dimensional
potential energy surface and does not encompass any ECs of the corresponding few-body systems,
it offers a valuable insight into the structure of the potential landscape. 
It is evident that the addition of particles to the system results in an increase not only of the mere
potential energy but in particular also of the complexity of the potential energy, attributable 
to the cumulative Coulomb interaction of the particles.
Of particular significance is the observation that the overall structure extends on what we have already been
finding for the two-body system: it exhibits (i) an overall oscillatory potential (ii) a hierarchical organization 
of energies and (iii) spatial irregularities in the minima/wells of the intersection potential.

Tentatively, the wells of the intersections (see Fig. \ref{fig:tbptombp}(b)) for small values
of $\Phi$ show a highly irregular oscillatory behaviour.
It is noteworthy that in general the potential wells become deeper upon transitioning to higher values of $N$. 
There exist however manifest exceptions. In Fig. \ref{fig:tbptombp}(b), this is at least visible
for the minima at $\Phi \approx 5/2$ and $\Phi \approx 7/2$. 
Increasing the number of particles for a given value of $\Phi$ (i.e. moving from the curve
for $N$ to the curve for $N+1$ in Fig. \ref{fig:tbptombp}(b))
results in a greater total extension $(\sum_{i=1}^{N-1} \Phi_i)$ of the particle chain.
Accordingly, the axial Coulomb forces exerted on the
outermost particles (the first and the last particle of the particle chain) prevail for $N+1$ particles
even at lower values of $\Phi$ compared to the $N$ particle chain. This ultimately leads to the flattening of the last potential well when $N$ is 
increased, as marked by the black crosses in Fig. \ref{fig:tbptombp}(b). Nonetheless, this attribute is 
not entirely generic, as in the transition from $N=2$ to $N=3$, two minima disappear, and for $6 \leq N \leq 8$, 
respectively $9 \leq N \leq 14$, the last minimum is stable. By increasing the geometric parameter $\rho$ of the helix, the last potential well
of each curve in Fig. \ref{fig:tbptombp}(b) would be shifted to smaller values of $\Phi$ and
the complex behavior of the intersection for $0 < \Phi < 1$ would vanish and become a
single potential well. This indicates a similar behavior of the many-body potential for an increase in $\rho$ compared to Fig. \ref{fig:tbptombp}(a). It is imperative to reiterate that Fig. \ref{fig:tbptombp}(b) is merely an intersection and does not embody the 
high-dimensional potential landscape. Nevertheless, this intersection partially reflects the structural peculiarities
of the potential landscape and its local equilibria, as we shall see in the subsequent sections.

\section{Methodology and computational approach}\label{chap:three}
The statistical and correlation analysis of the spatial and energetic
properties of the ECs of a long-range interacting
particle system on a helix necessitates the finding of all local minima of our high-dimensional
oscillatory potential energy surface (see Eq. \eqref{eq:hamiltonianinternal}).
This is illustrated in Fig. \ref{fig:tbptombp}(b) and will be further elaborated
below: the potential energy landscape shows a mix of regular and irregular
oscillations and undergoes substantial fluctuations across various particle
configurations while maintaining a hierarchical structure. The exponential growth in configuration space volume,
concomitant with the rise in dimensionality, poses a major challenge in identifying all local
equilibria of the potential landscape. We address this by implementing a multi-start approach,
entailing the definition of multiple sampling points. Each sampling point is then utilized
as an initial condition for a local optimizer. Notwithstanding the aforementioned "disorder" inherent
in the potential landscape, the repulsion within and across helical windings must engender approximate
regularity in the occurrence of ECs, which motivates the definition of the sampling points as a periodic
hyperlattice $L \subseteq Q$ in the configuration space $Q \subseteq \R^{N-1}$. Every point in this 
space is given by a vector $\mathbf{\Phi} = (\Phi_1, \Phi_2, \ldots, \Phi_{N-1})$ of the internal
coordinates. The hyperlattice is defined as 
$L = \tilde{\Phi}_1 \times \tilde{\Phi}_2 \times \ldots \times \tilde{\Phi}_{N-1}$, where the subsets
$\tilde{\Phi}_k \subseteq \R$, with $k = 1, 2, \ldots, N-1$, define $n$ evenly spaced points in the 
$k$-th dimension in a given interval $[a,b] \subseteq \R$. 
While $a=0.01$ was fixed, $b$ was chosen in the range $4 \leq b \leq 10$, depending on $N$.
The optimization algorithm employed is
\textit{L-BFGS-B} \cite{liuLimitedMemoryBFGS1989,byrdLimitedMemoryAlgorithm1995}, which implements
a quasi-Newton method with bound constraints for the $k$-th coordinate.

The computational effort resulting from this approach is increasing exponentially with the number
of particles employed in the system ($\propto n^{N-1}$). Furthermore, the highly oscillatory
structure of the potential landscape requires a high sampling point density
$\delta = n/(b-a)$ to find all local equilibria. Reducing the search space is
one way to lower the computing costs under these conditions. As indicated in the previous sections,
the maximum value of $\Phi_k$ that allows for ECs either remains approximately constant or decreases
when a particle is added to the chain. Consequently, this maximum value serves as an upper
bound constraint $b$ for the system with an incrementally larger number of particles.
On the other hand, the inversion of the helix with a given configuration 
$\mathbf{\Phi}$ results in an identical configuration $-\mathbf{\Phi}$ with the same potential energy. Therefore, the search
space can be further reduced by approximately half through the elimination of these identical configurations. We accomplish this
by employing a recursive constraint for $\Phi_i$ and $\Phi_{N-i}$, $i=1,\ldots,k=\lfloor (N-1)/2 \rfloor$
 and $N \geq 3$. 
We enforce that $\Phi_1 \geq \Phi_{N-1}$; $\Phi_2 \geq \Phi_{N-2}$ if $\Phi_1 = \Phi_{N-1}$;
$\Phi_3 \geq \Phi_{N-3}$ if $\Phi_1 = \Phi_{N-1} \land \Phi_2 = \Phi_{N-2}$; $\ldots$ ;
$\Phi_{k} \geq \Phi_{N-k}$ if
$\Phi_1 = \Phi_{N-1} \land\, \ldots \,\land \Phi_{k-1} = \Phi_{N - (k-1)}$.
Formally, this constraint can be
written as a lexicographical ordering of the subsets $A = \lrbrace{\Phi_{1},\ldots,\Phi_{k}}$
and $B = \lrbrace{\Phi_{N-1},\ldots,\Phi_{N-k}}$: $A \geq_{\mathrm{lex}} B$.
In our optimization algorithm, points of $L$ that invalidate this constraint 
are introduced into the interstitial voids of the lattice structure.

The chosen algorithm reliably converges to local equilibria in domains characterized by a high degree
of non-linearity and oscillatory behavior in the potential landscape. However, the presence of flat
minima for large interparticle distances and the overall decreasing behavior of the potential
energy result in a significant number of sampling points whose optimization is terminated
due to reaching the upper bound constraint. Therefore, it is imperative that the density
$\delta$ is sufficiently high in order to determine all flat minima and to ensure the
coverage of the highly oscillatory structure for low distances. For $2 \leq N \leq 7$,
approximate completeness was ensured by conducting multiple iterations with increasing $\delta$ until the 
obtained number of ECs attained convergence at a specified value. The computing time for
$2 \leq N \leq 4$ was relatively low, so starting with $\delta \approx 10$ directly reached
convergence. For $5 \leq N \leq 7$, convergence was attained at $\delta \approx 12$.
Due to the disproportionate computational costs associated with studying the convergence
for $N=8$, a lower density of sampling points ($\delta \approx 10$) was employed, based on the 
minimum spatial distance of the ECs for $N=7$, while the search space for each sampling
point was limited to a volume up to the next grid neighbors with an added overlap, thereby
reducing redundancy. However, completeness is not guaranteed and probably not reached
in this case, but the potentially missed ECs represent a very small fraction of their total
number.

Upon initial observation, the employed approach may appear to be suboptimal. Even though multi-start approaches
are common, they typically employ random sampling or quasi-random sampling methods, such as
Sobol sequences or Latin Hypercube Sampling. Nonetheless, the repulsion within a winding between
the particles leads to an underlying grid-like pattern that is more efficiently covered by a dense
periodic hyperlattice. While the high discrepancy inherent in random sampling is not a viable option,
quasi-random sampling methods have the potential to demonstrate comparable efficacy. 
Additionally, the number of ECs attained through the chosen approach is disproportionate to the number
of initial guesses. The majority of the obtained results were terminated due to the optimization
reaching a bound constraint or because the results were duplicates of previously determined ECs.
A reduction of redundancy may be achieved through the implementation of an approach analogous to
that employed in the context of eight particles, or by penalizing ECs that have already been identified.
Moreover, the elimination of subspaces devoid of ECs can result in a substantial enhancement in
computational efficiency. Nevertheless, this approach is challenging due to the overall decreasing
potential energy. However, the implementation of certain spatial constraints can be informed by
the empiricism of the ECs gathered thus far.


\section{Results and Discussion}\label{chap:four}
The subsequent section will offer a statistical and correlational analysis of the ECs of the long-range interacting
particle system on the helix. In order to study the system in a regime that provides ECs between regularity
and disorder, we have selected $\rho = 0.8$. This yields generic statistical results for this
particular regime. Initially, the spatial characteristics will be addressed through two primary metrics:
the totality of interparticle distances present in the ECs and the distribution of the ECs in the 
high-dimensional configuration space $Q$. Subsequently, an investigation of the energies corresponding to the ECs will be 
conducted. Ultimately, the correlations between the energetic and spatial properties will be examined.

\subsection{Spatial equilibrium configurations: Statistics and correlations}
\begin{figure}[t]
    \centering
    \includegraphics[width=\columnwidth]{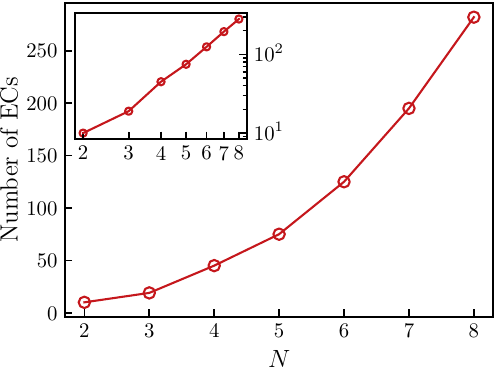}
    \caption{The total number of attained ECs, contingent upon the number of particles. The inset illustrates the number
             of ECs by $N$ on a double logarithmic scale, exhibiting that it follows a power law $(\approx N^{2.7})$.}
    \label{fig:number_minima}
\end{figure}
The analysis of the statistics and correlations of the spatial properties of the ECs will be conducted
for each particle number in the range 2-8.
In consideration of the denotation of an EC as $\mathbf{\Phi}_0 = (\Phi_0^1, \Phi_0^2, \ldots, \Phi_0^{N-1})$,
the distribution of the totality of the interparticle distances $\Phi_0^i$, $i = 1,2, \ldots, N-1$ of all 
ECs will be examined, independent of their sequential ordering within an EC. Building upon this distribution, a 
mapping of these $\Phi_0^i$ to distinct clusters will be employed to elucidate the spatial structure of the ECs.
This will be achieved by indexing the clusters and then correlating the $\Phi_0^i$ values within an EC based
on their assigned cluster index.
Finally, we will gain insight into the spatial structure by studying the distribution of ECs within
the $(N-1)$-dimensional potential landscape, quantifying their Euclidean distances in the configuration
space.

Nonetheless, prior to the examination of the spatial characteristics of the ECs, their quantity in relation
to the number of particles will be considered. This is shown in Fig. \ref{fig:number_minima} for $2 \leq N \leq 8$.
The increase in complexity and
dimensionality of the potential landscape results in a rapid growth in local equilibria as $N$ increases.
For $N=2$, the system exhibits 10 ECs, while for $N=8$, 282 ECs exist. Despite the disproportion between the increase 
in configuration space volume and the increase of ECs by $N$, the latter exhibits a power-law behavior $(\approx N^{2.7})$, as evidenced 
by the inset of Fig. \ref{fig:number_minima}, presenting the number of ECs by $N$ on a double logarithmic scale.
Nevertheless, this assertion is applicable only in the local context for $N = 2,\ldots,8$,
due to the insufficient statistical significance to draw conclusions about the global behavior of the number of ECs for larger $N$.
This uncertainty is further underpinned by the fact, that 
for transitioning from $N=2$ upon $N=4$ the addition of a particle each 
leads to roughly doubling the number of ECs, thereby indicating exponential growth. Albeit, this behavior diminishes for $N > 4$. 
However, the subsequent spatial analysis might explain some of the underlying causes of the observed cessation of 
exponential growth.

\subsubsection{Distribution of interparticle distances}
\begin{figure}[t]
    \centering
    \includegraphics[width=\columnwidth]{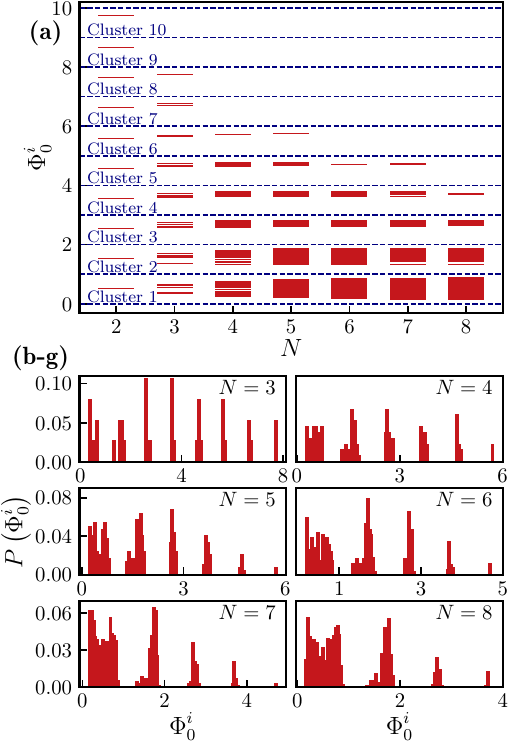}
    \caption{\textbf{(a)} Distribution of interparticle distances $\Phi_0^i$ of all ECs by the number of particles $N$. Each finite,
             red, horizontal line corresponds to one $\Phi_0^i$. Note that some values of $\Phi_0^i$ can
             occur multiple times within an EC or across different ECs, condensing to one horizontal line.
             Furthermore, similar $\Phi_0^i$ can lead
             to a high density of lines, resulting in an
             almost continuous range of $\Phi_0^i$ values.
             The occurring values clusterize, where the clusters are
             indexed as well as subdivided by the blue dashed lines. \textbf{(b-g)} Histogram of the values $\Phi_0^i$ showing
             the probability $P(\Phi_0^i)$ of their occurrence within defined intervals.}
    \label{fig:cluster_plus_histogram}
\end{figure}
\begin{figure*}[ht]
    \centering
    \includegraphics[width=\textwidth]{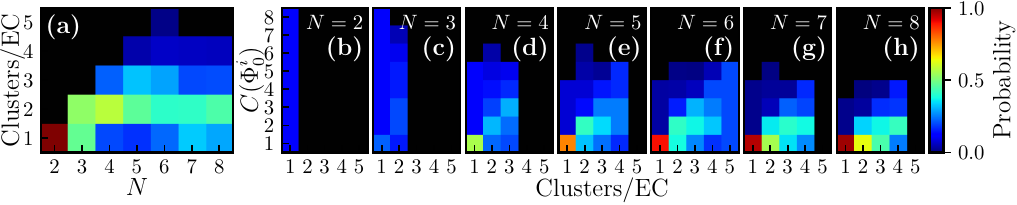}
    \caption{\textbf{(a)} The probability on how many distinct clusters the interparticle distances $\Phi_0^i$ 
             of an EC (Clusters/EC) are distributed in dependence of the number of particles $N$. 
             For example, for the EC $\mathbf{\Phi}_0 \approx (1.35, 0.36)$,
             $C(\Phi_0^1) = 2$ and $C(\Phi_0^2) = 1$, so $\Phi_0^i$ is distributed on two Clusters in that EC.
             The probability is normalized for each $N$.
             \textbf{(b-h)} The probability on which specific clusters the $\Phi_0^i$ are distributed, given the
             value of Clusters/EC. Note that $C(\Phi_0^i)$ is cut at the value of 8 because $C(\Phi_0^i) = 9,10$ are
             only relevant for the trivial case of $N=2$. The probability is normalized for each value of Clusters/EC.}
    \label{fig:num_clusters_ec}
\end{figure*}
As indicated in Fig. \ref{fig:tbptombp}(a), the underlying regularity in the occurrence of ECs of the helical two-body
system must also be reflected in the potential landscape of the few-body system, particularly in the occurring
interparticle distances as a result of the repulsion between two particles within a helical winding.
In fact, the distribution of all interparticle distances $\Phi_0^i$ of the ECs still represents the underlying
regularity, as illustrated in Fig. \ref{fig:cluster_plus_histogram}(a). It demonstrates that
the distances $\Phi_0^i$ can be assigned into distinct clusters, where distances of multiples of one
helical winding $(0,1,2,\ldots)$ can be utilized to separate the clusters. By implementing an ascending indexing
scheme, each $\Phi_0^i$ can be assigned to the cluster $C(\Phi_0^i) = \lceil \Phi_0^i \rceil$.

A particularly intriguing aspect is the crossover from $N=2$ to higher particle numbers.
Due to the repulsive long-range interactions, the highest clusters in each case - and 
therefore the occurrence of such large interparticle distances - disappears completely when a particle is
added to the system: This can be seen in the transition where $N$ is 2-3-4, 5-6 and 7-8.
Upon transitioning from $N=2$ to larger $N$, the clusters 3-6 undergo an initial expansion in width
into the direction of greater values of $\Phi_0^i$, until a further increase in $N$ prompts a reversal
in transitional behavior, resulting in a decrease in cluster width in both directions to greater and
smaller values of $\Phi_0^i$. This process continues until a single value 
of $\Phi_0^i$ is occupied, which subsequently vanishes upon the addition of another particle.
This is further elucidated in Fig. \ref{fig:cluster_plus_histogram}(b-g), which illustrates
the distribution of $\Phi_0^i$ by showing the probability of their occurrence within defined intervals.
Herein, the described behavior is reflected
in a transition of the corresponding probability distributions from asymmetric to symmetric by an increase
in $N$. 

A generally similar but more complex transition can be observed for the Clusters 1 and 2 (see Fig. \ref{fig:cluster_plus_histogram}(a-g)).
While the maximum value of $\Phi_0^i$ in these clusters also initially increases, the minimum value
of $\Phi_0^i$ is decreasing, thereby forming a fine structure with two subclusters. For Cluster 2
the upper subcluster is broader than the lower one, evidenced by an asymmetry in the corresponding probability
distributions. For $N>5$, cluster 2 decreases, predominantly by the fading of the lower subcluster. 
In contrast, cluster 1 exhibits a less pronounced dominance of the upper subcluster, characterized by
a more equitable distribution of $\Phi_0^i$ across both clusters. This results in probability distributions,
that are more or less symmetrical. Additionally, cluster 1 merely shows an increase
in its width, while a subsequent decrease may only commence at higher values of $N$.

The distribution under consideration reveals the repercussions of repulsive long-range interactions, which hinder
the formation of stable ECs over extended interparticle distances within the few-body system, demonstrated
by a constant decrease of the maximum value of $\Phi_0^i$ as $N$ increases.
This is accompanied by a growth in the occurrence of small interparticle distances
in the ECs with an increase in $N$, particularly evident in cluster 1, where a high probability
for interparticle distances of even less than half a helical winding emerges. 
Additionally, a transition from the highly regular two-body system to an increasing
degree of disorder is observed within the growth of the addressed clusters. This thereby results in
a higher variety of possible interparticle distances. Nevertheless, the clusterization itself
reflects the underlying regularity.

\subsubsection{Cluster correlations}
The clusterization of $\Phi_0^i$ can be further used as a metric to understand the inherent structure of
sequences $\mathbf{\Phi}_0 = (\Phi_0^1, \Phi_0^2, \ldots, \Phi_0^{N-1})$. Therefore,
Fig. \ref{fig:num_clusters_ec}(a) shows the probability on how many distinct clusters the $\Phi_0^i$ of an EC
are distributed (Clusters/EC or number of clusters per EC) for $N$ in the range of 2-8. 
Evidently, the number of internal coordinates per EC 
($N-1$) serve as an upper limit for the number of clusters per EC. This results in a non-zero probability of
an incremental number of clusters per EC with each increase in $N$ for $2 \leq N \leq 6$.
However, these probabilities are decreasing as $N$ increases.
This is due to the instability of
configurations with large interparticle distances as $N$ increases, as well as the simultaneous increase
in configurations with low interparticle distances. This ultimately results in clusters whose widths are
uneven, thereby allowing for fewer combinations where all $\Phi_0^i$ are in different clusters within an EC.
Finally, for $N > 6$ this behavior is interrupted as for $N=7$, the occupation 
of 5 clusters per EC would be theoretically possible but does not occur. For $N=8$ the maximum occuring number 
of Clusters/EC is then equal to the total number clusters (see Fig. \ref{fig:cluster_plus_histogram}(a-g)).
Apart from the trivial two-body case, a high probability for two clusters per EC
is observed for all $N$. For $N=3$ a similarly high probability occurs for one cluster per EC while
for $4 \leq N \leq 6$, there is a higher probability 
of more than two clusters per EC than of just one. The opposite is true for $N=7,8$, which simply follows
from the large number of low interparticle distances that occupy the first and second clusters
(see Fig. \ref{fig:cluster_plus_histogram}).

\begin{figure*}[t]
    \centering
    \includegraphics[width=\textwidth]{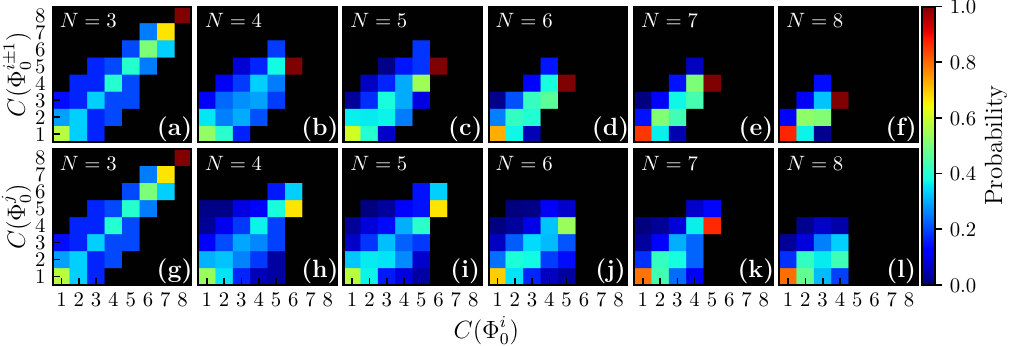}
    \caption{\textbf{(a-f)} The conditional probability (normalized for each $C(\Phi_0^i)$) that the next neighboring distance $\Phi_0^{i\pm 1}$ of $\Phi_0^i$
             is in the cluster $C(\Phi_0^{i\pm 1})$, if $\Phi_0^i$ is in the cluster $C(\Phi_0^i)$ for $3 \leq N \leq 8$. 
             \textbf{(g-l)} The conditional probability (normalized for each $C(\Phi_0^i)$) that an arbitrary other interparticle distance $\Phi_0^j$ than $\Phi_0^i$
             (so $i\neq j$) of an EC is in the cluster $C(\Phi_0^{j})$ if $\Phi_0^i$ is in the cluster $C(\Phi_0^i)$,
             for $3 \leq N \leq 8$.}
    \label{fig:cluster_correlation}
\end{figure*}

In order to comprehend not only the total quantity of clusters per EC but also the
probability distribution of the interparticle distances on the specific clusters,
Fig. \ref{fig:num_clusters_ec}(b-h) illustrates the probability
of the occurence of the cluster $C(\Phi_0^i)$ within an EC, contingent on the number of clusters per EC.
A discernible transition from an even to a highly uneven distribution is evident as the crossover progresses 
from $N = 2$ to $N = 8$.
It is noteworthy that the probability of $C(\Phi_0^i) = 1$ exhibits a remarkable upward trend as $N$
increases, in the event that also the Clusters/EC are equal to one. For $N=8$, this probability
approaches one.

In a similar vein, as the number of clusters per EC increases, a transition to higher probabilities for lower 
values of $C(\Phi_0^i)$ becomes evident, commencing from $N = 4$, where the probabilities on the diagonal 
prevail, to $N = 8$. It therefore demonstrates, that if the number of clusters per EC is high, intermediate
interparticle distances, assignable to the clusters 2 or 3 are more probable than those
that are remarkably low or high. However, the probability of low distances is generally increasing with $N$, 
due to the sheer increase in their quantity. Based on the last mentioned and
the repulsive long-range interactions, interparticle distances with $C(\Phi_0^i)=1$ are likely if the
number of clusters per EC is low.

Obtaining information about the ordering of $\Phi_0^i$ within the ECs is the natural subsequent step.
Therefore, Fig. \ref{fig:cluster_correlation}(a-f) shows the probability that the next neighboring
distance in an EC, $\Phi_0^{i\pm 1}$, of $\Phi_0^i$ is assignable to cluster $C(\Phi_0^{i\pm 1})$,
given that $\Phi_0^i$ is in $C(\Phi_0^i)$.
For all $N$, a diagonally dominated probability
distribution is observed, indicating that the next neighboring distances of $\Phi_0^i$ are, with a high
probability, in the same or in a neighboring cluster as $\Phi_0^i$ itself. However, the probability that 
one of the next neighboring distances is in a lower cluster than $\Phi_0^i$ ($C(\Phi_0^{i\pm 1}) < C(\Phi_0^i)$)
is slightly higher than being in a higher
cluster. This behavior is predominantly exhibited when $\Phi_0^i$ is situated within the uppermost
cluster for $4 \leq N \leq 8$. In this case, $\Phi_0^{i\pm 1}$ is consistently situated within the lower
neighboring cluster. For $N=3$, this does not apply, given that $C(\Phi_0^{i\pm 1}) = C(\Phi_0^i)$, if 
$\Phi_0^i$ occupies the highest possible cluster. 
It is noteworthy, that the probability of $C(\Phi_0^i) = C(\Phi_0^{i\pm 1}) = 1$
is notably high and exhibits a marked increase upon transitioning from $N=3$ to $N=8$.

Fig. \ref{fig:cluster_correlation}(g-l) extends this correlation to
the whole sequence $\mathbf{\Phi}_0$ of an EC. Therefore, the $y$-axis now shows the probability that 
an arbitrary distance $\Phi_0^j$ of an EC, is in
the cluster $C(\Phi_0^j)$, for $i \neq j$. The diagonal dominance, demonstrated in
Fig. \ref{fig:cluster_correlation}(a-f), appears to be slightly smeared and the probability of finding
$\Phi_0^j$ in a lower cluster than $\Phi_0^i$ is even more pronounced.
However, when $\Phi_0^i$ is in the cluster 2,3 or 4, a reasonably wide distribution
for the probability of the cluster of $\Phi_0^j$ is observed. Concurrently, the distributions for $C(\Phi_0^j)$,
if $\Phi_0^i$ is in the first or the highest cluster are highly uneven. Nonetheless, they exhibit a higher probability
of $\Phi_0^j$ being in a higher cluster compared to $\Phi_0^{i\pm 1}$ in the corresponding case. 
However, this does not apply for $N=6$ and $N=8$.

In summary, the correlations of the corresponding cluster indices of $\Phi_0^i$ and $\Phi_0^{i\pm 1}$, respectively
$\Phi_0^j$, are analogous, and collectively demonstrate that similar values of $\Phi_0^i$ occur together in ECs.
Nonetheless, for interparticle distances that are assignable to intermediate clusters (2,3,4), the distribution of the 
interparticle distances that occur together with them in ECs is reasonably wide and shifted to lower clusters.
It allows for the conclusion that the predominant repulsion in the axial direction for the discussed few-body system on the helix
primarily allows for ECs with high to intermediate interparticle distances, provided that these distances are also
accompanied by smaller interparticle distances for the other particles. 
Additionally, in conjunction with the strong correlation between proximate clusters for next neighboring
interparticle distances, it is indicated that markedly disparate interparticle distances are unlikely to occur
as neighboring distances in an EC. Rather, it is more likely that the interparticle distances within the
sequence of $\mathbf{\Phi}_0$ are hierarchically ordered.
Non-trivially, configurations with high interparticle distances (which are infrequent in general) are
more likely to be predominantly comprised of interparticle distances that fall within the same
or neighboring lower cluster (see $N=3,4,5,7$) due to the long-range repulsion.
Finally, the elevated probability that low interparticle distances co-occur within an EC,
is also a consequence of the predominance of the first cluster of $\Phi_0^i$ (see Fig. \ref{fig:cluster_plus_histogram}(a)).

\begin{figure}[t]
    \centering
    \includegraphics[width=\columnwidth]{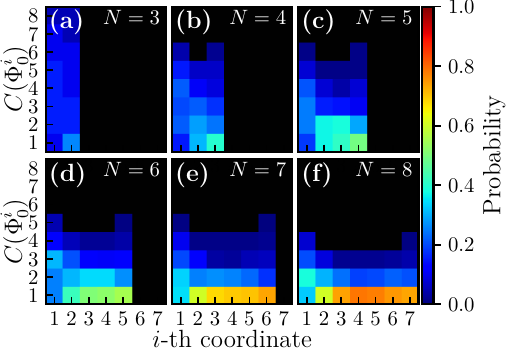}
    \caption{The probability for $\Phi_0^i$ being in Cluster $C(\Phi_0^i)$ in dependence of the index $i$, i.e,
    the $i$-th coordinate for $3 \leq N \leq 8$. The probability is normalized for each $i$-th coordinate.}
    \label{fig:pos_i_cluster}
\end{figure}

To gain full statistical insight into the ordering of interparticle distances within an EC,
Fig. \ref{fig:pos_i_cluster}(a-f) further provides positional information, by showing the probability
of the cluster value $C(\Phi_0^i)$, contingent on the index $i$, i.e., the $i$-th coordinate of an EC.
A general trend that emerges from the transition from $N=3$ to $N=8$, is the shift in the probability distributions
for $C(\Phi_0^i)$ from nearly even to highly uneven distributions. However, the distribution for the first coordinate
remains relatively even across all values of $N$.
In contrast, the illustration reveals a consistently high and increasing probability that the $(N-1)$-th distance
falls within the first cluster as $N$ increases. However, a more intriguing observation is that, with an increase in
$N$, the probability that the interparticle distances between the inner particles ($i = 2,\ldots,N-2$) exhibit a 
stronger increasing probability of being in cluster 1. Concurrently, for $4 \leq N \leq 8$, the highest
occupiable cluster is exclusively reached by the first ($i=1$) and the last coordinate ($i=N-1$).
Therefore, stable ECs occur with a high probability in circumstances where the inner particles are in closer proximity
to one another than the outermost particles are to their neighbored particle. 
Concomitantly, Fig. \ref{fig:pos_i_cluster}(a-f) reveals a weak symmetry between the probabilities of $C(\Phi_0^i)$ and $C(\Phi_0^{N-i})$,
indicating an inherent symmetry between the values of $\Phi_0^i$ and $\Phi_0^{N-i}$ in at least
some ECs. Separately, an overall high probability for a hierarchy of the values $\Phi_0^i$
within the sequence of an EC from high interparticle distances for $i=1$ to smaller ones for $i=N-1$,
is evident. In summary, there is a high probability for ECs to be either symmetric in $\Phi_0^i$
and $\Phi_0^{N-i}$ or hierarchically ordered in $\Phi_0^i$.

\subsubsection{Helical configurations}
\begin{figure}[t]
    \centering
    \includegraphics[width=\columnwidth]{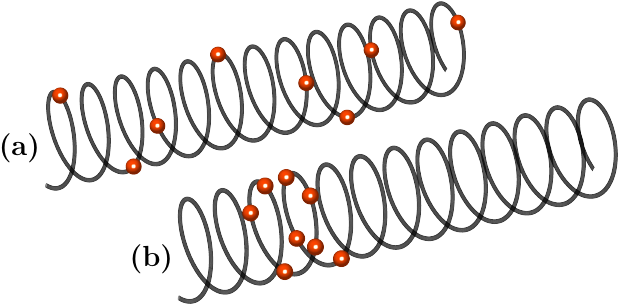}
    \caption{Helical equilibrium configurations for $N=8$ with 
    \textbf{(a)} $\mathbf{\Phi}_0 = (1.64, 1.72, 1.73, 1.73, 1.73, 1.72, 1.64)$ and 
    \textbf{(b)} $\mathbf{\Phi}_0 = (0.26, 0.25, 0.23, 0.17, 0.23, 0.25, 0.26)$.}
    \label{fig:helical_configurations}
\end{figure}
\begin{figure*}[t]
    \centering
    \includegraphics[width=\textwidth]{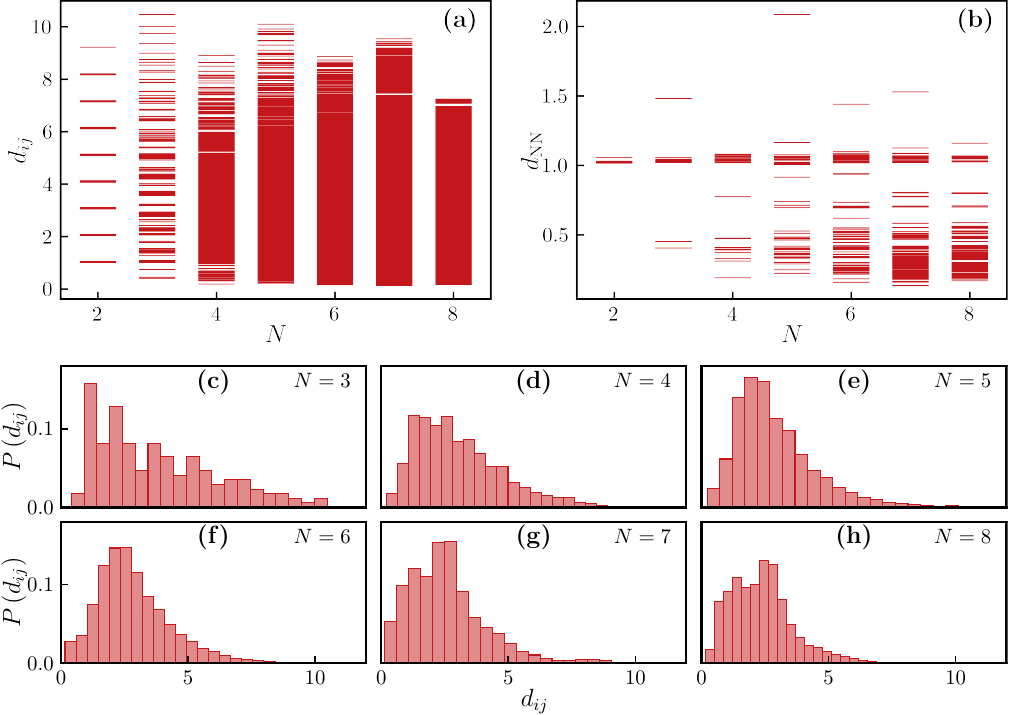}
    \caption{\textbf{(a)} Distribution of Euclidean distances $d_{ij} = \lvert\lvert\mathbf{\Phi}_0^i - \mathbf{\Phi}_0^j\rvert\rvert$
             between all ECs for $i \neq j$ in the configuration space $Q$ and for different $N$. Each red line corresponds to a pairwise distance.
             \textbf{(b)} Distribution of distances of nearest-neighbor ECs
             $d_{\mathrm{NN}} = \lvert\lvert\mathbf{\Phi}_0^i - \mathbf{\Phi}_0^j\rvert\rvert$ of $\mathbf{\Phi}_0^i$, i.e., for 
             $\lvert\lvert\mathbf{\Phi}_0^i - \mathbf{\Phi}_0^j\rvert\rvert \leq \lvert\lvert\mathbf{\Phi}_0^i - \mathbf{\Phi}_0^k\rvert\rvert$ for $i\neq j \neq k$,
             in the configuration space $Q$. For (a) and (b) each finite, red, horizontal line corresponds to
             one distance $d_{ij}$, respectively $d_{\mathrm{NN}}$. Multiple identical values can appear as one
             line. The high density of occurring values leads to an almost continuous range of distances.
             \textbf{(c-h)} Histogram of the distribution of distances $d_{ij}$, which shows the probability
             $P(d_{ij})$ for the occurrence of a distance $d_{ij}$ in a defined interval.}
    \label{fig:distances_configuration_space}
\end{figure*}

In addition to the statistical analysis of interparticle distances, the occurrence of intrinsically helical ECs should be emphasized. This refers to ECs,
in which the particle chain exhibits a helical configuration due to its inherent particle ordering. The assumption that an EC is helical is made under the
following criteria for the $\Phi_0^i$ of an EC: $\Phi_0^i \neq 1$ and $\Phi_0^1 \approx \Phi_0^2 \approx \ldots \approx \Phi_0^{N-1}$, where
deviations from exact equidistancy such that $\Phi_0^{i,\mathrm{max}} - \Phi_0^{i,\mathrm{min}} < 0.2$ are tolerated. Therefore, these ECs can be roughly
associated with the potential wells of the global diagonal of the potential landscape (see Fig. \ref{fig:tbptombp}(b)). Such configurations
are illustrated in Fig. $\ref{fig:helical_configurations}$ for $N=8$ and for $\Phi_0^i$ within the second (a) and within the first (b) cluster.
The existence of stable ECs in the form of a helical geometry underscores its significance in the context of stable configurations of repulsive long-range interacting
particle systems, not only in the case of helical confinement but also for analogous systems with cylindrical symmetric confinement
(see Refs. \cite{tsytovichHelicalStructuresComplex2005b,tsytovichHelicalStructuresComplex2005a,kamimuraCoulombDoubleHelical2012}).

\subsubsection{Euclidean distances of ECs in the configuration space}
The gained understanding of the ECs, employed by correlations of the 1D metric of the interparticle distances,
does not appropriately reflect the structure of the ECs $\mathbf{\Phi}_0 \in Q$ within the high dimensional potential landscape.
Despite the challenges associated with the acquisition of structured information from high-dimensional data,
it is possible to examine the relational information between the data points. This includes
the pairwise Euclidean distances 
$d_{ij} = \lvert\lvert\mathbf{\Phi}_0^i - \mathbf{\Phi}_0^j\rvert\rvert$, $i \neq j$ between the ECs in the configuration
space $Q$. The distribution of these pairwise distances is shown in Fig. \ref{fig:distances_configuration_space}(a) for
different $N$. Upon initial observation, it is evident that there is a transition from approximately equidistant
$d_{ij}$ for $N=2$ to the emergence of an increasing continuous interval in the corresponding distributions
as $N$ increases to a nearly totally continuous distribution for $N=8$.
The three-body case can be considered as
an intermediate state between equidistancy and continuity, thereby appearing totally irregular.
Given the rapid increase in total quantity of pairwise distances with the number of ECs ($(n-1)\frac{n}{2}$),
and the concomitant rise in dimensionality that engenders further possible distances, 
the exhibited behavior can be interpreted as a sort of density effect.
Nevertheless, this line of argumentation is not entirely compelling. In the context of a fully periodic hyperlattice,
the pairwise distances would still be staggered, bearing a resemblance to the behavior observed in the two-body system.
Thus, the transition to continuity can be interpreted
as a rise in complexity and irregularity in the occurrence of the ECs, which results in a variety of different 
distances with small deviations, thereby forming a continous range of occurring pairwise distances $d_{ij}$.

With respect to Fig. \ref{fig:tbptombp}(b), the observed behavior is likely attributable to a strongly
oscillatory potential landscape, particularly for small interparticle distances.
Supporting this, the growth of the continuous intervals of $d_{ij}$ as $N$ increases is
accompanied by a significant increase in the probability that $\Phi_0^i$ is in the first cluster,
as substantiated in the preceding analysis.
Furthermore, in Fig. \ref{fig:distances_configuration_space}(c-h), showing histograms of the concerning distribution,
a rapid increase in the probability for small to intermediate $d_{ij}$ with a suppressed tail
for large $d_{ij}$, as $N$ is increased, can be observed. Although it is primarily the small interparticle distances that constitute
the ECs for higher $N$, it has to be noted that edge effects, in conjunction with the pairwise evaluation,
lead to a shift to a high quantity of intermediate $d_{ij}$.
Additionally, the oscillation in
the maximum distance of $d_{ij}$ (see Fig. \ref{fig:distances_configuration_space}(a)) by the alteration of $N$ is noteworthy.
An increase in the maximum distance can be simply explained by an increase in dimensionality.
Conversely, decreases in the maximum distance $d_{ij}$ occur when the maximum value of $\Phi_0^i$
in the ECs decreases by approximately one helical winding (see Fig.
\ref{fig:cluster_plus_histogram}(a)). Ultimately, the minimum value of $d_{ij}$ appears to be converging
against a certain value, which can be attributed to a predominant repulsion at a certain cutoff value.

The hypothesis that the increasing variety in small interparticle distances is part of the forming continuity is further
supported by Fig. \ref{fig:distances_configuration_space}(b), which shows the distance $d_{\mathrm{NN}}$ of an
EC to its nearest neighbored EC in the configuration space $Q$. One can observe the formation of two main
clusters at $d_{\mathrm{NN}} \approx 1$ and for $0.1 \lessapprox d_{\mathrm{NN}} \lessapprox 0.9$, whereby the latter emerges
for $N=3$. While the upper cluster at $d_{\mathrm{NN}} \approx 1$ initially increases and then
decreases in width, as $N$ increases, it exhibits a relatively high degree of regularity, i.e., a low width,
in comparison to the other cluster. The latter exhibits a high cluster width, whereas for $N \geq 4$ the level
density increases, forming continuous intervals and thereby a high variety of small
nearest-neighbor distances $d_{\mathrm{NN}}$. Therefore, it can be concluded that not just the intersection
for equidistant particles (see Fig. \ref{fig:tbptombp}(b)), but also the high-dimensional potential landscape
exhibits a complex oscillatory behavior for small interparticle distances.
Furthermore, there exist also nearest-neighbor distances for values at $d_{\mathrm{NN}} \approx 1.5$ and
$d_{\mathrm{NN}} \approx 2$. They most likely correspond to the outermost EC (the EC with the largest total extension of the particle chain),
which appear to be 
somewhat isolated. However, these values of $d_{\mathrm{NN}}$ are statistically insignificant
and it is possible that not all ECs have been identified in this particular 
instances.

In summary, the spatial properties were initially demonstrated to exhibit an increase in disorder with an
increase in the number of particles, which is evident in the distribution of interparticle distances
of the ECs, which form clusters of interparticle distances around the regular distance for $N=2$.
This transition to increased disorder can also be observed in the distribution of distances
of ECs in the configuration space $Q$. As $N$ increases, the distribution of distances becomes
continuous, thereby implying a highly complex oscillatory behavior of the potential landscape for small
interparticle distances.
However, the clusterization of the interparticle distances still contains regularity that is also reflected
in a strong correlation of the $\Phi_0^i$ within an EC, based on the cluster metric.
It has been demonstrated that next neighboring interparticle distances in ECs
likely are in neighboring clusters and that the ECs in general
tend to exist with a hierarchy in $\Phi_0^i$ or a symmetry in $\Phi_0^i$ and $\Phi_0^{N-i}$.
Ultimately, it is significant
that the inner particles particles exhibit a tendency to be in closer proximity to each other than the outermost particles
to their neighbored particle.

\subsection{Energetical properties of equilibria}
\begin{figure}[t]
    \centering
    \includegraphics[width=\columnwidth]{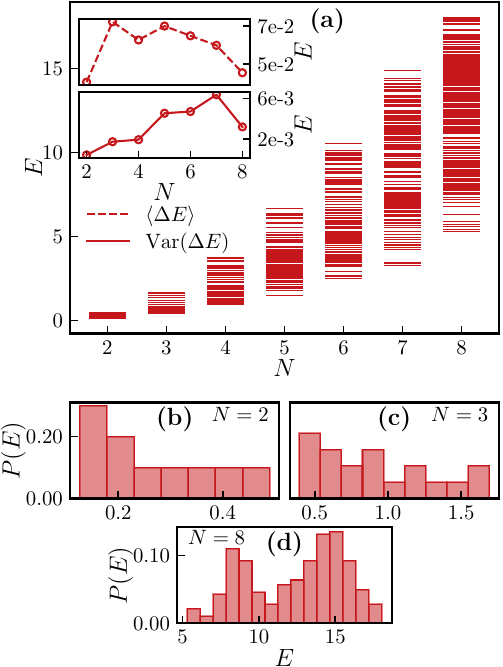}
    \caption{\textbf{(a)} The distribution of the potential energies $E \equiv V$, corresponding to the ECs found for $N = 2,\ldots,8$. The insets
             show the mean spacing $\langle \Delta E \rangle$ of the energies as well as the corresponding
             variance $\mathrm{Var}(\Delta E)$.
             \textbf{(b-d)} Histogram, showing the probability $P(E)$ of finding an energy within a defined interval of $E$, for
             $N=2,3,8$.}
    \label{fig:energy_distribution}
\end{figure}
The subsequent part will undertake a detailed examination of the potential energies associated with the ECs that have
been previously analyzed.
Fig. \ref{fig:energy_distribution}(a) illustrates the distribution of potential energies $E\equiv V$ for $N = 2-8$.
Each illustrated line corresponds to the energy of an EC and is referred to as an energy level.
In accordance with the cumulative potential energy term a trend of increasing energy levels as $N$ increases
can be identified.
Nonetheless, the overlap of energy ranges for systems with different N, as well as the total energy 
range, increases with increasing $N$.
However, the number of ECs is growing faster than the total energy range, consequently leading
to an overall decrease in the mean spacing $\langle \Delta E \rangle$ of the energy levels.
This is shown in the upper inset. Contrary to the prevailing trend, an increase in the mean spacing is observed in the 
transitions from $N = 2$ to $N = 3$ and from $N = 4$ to $N = 5$.
Especially the former transition differs significantly; however, it represents an outlier in the statistical analysis of the few-body system.

Furthermore, as $N$ increases, the energy levels exhibit a higher variance in their distribution,
manifested in energy ranges that exhibit significantly high level densities.
As demonstrated in the lower inset, the increase in energy spacing variance is weak for $N=4,6$ and a
subsequent decrease for $N=8$ occurs, where the latter is reflected in a broad energy range with a high level density.
The increase in these high level density ranges exhibits a structural transition
as $N$ increases: a shift from low energies for $N=2,3$ to intermediate energies for $N \geq 4$, and finally,
the emergence of a second, separate range with dense energy levels for large values of $E$ for $N \geq 7$.
In Fig. \ref{fig:energy_distribution}(b-d) this transition is exemplified by the probability
of locating an energy level within a defined interval of $E$. 
Indicated by a heightened probability for intermediate energies for $N=3$ in comparison to $N=2$, the
mentioned shift to intermediate energies already commences at $N=3$.
For $N=8$, two distinct maxima of energies within a defined interval are discernible.
The distribution around the second maximum,
which corresponds to high energies, is broader and exhibits a higher probability than the first.

The substantial increase in low interparticle distances (see Fig. \ref{fig:cluster_plus_histogram}(a-g))
for the ECs for $N=7,8$ is indicative of a secondary elevated energy range with a high level density, suggesting
that the energies in that range correspond to ECs that predominantly exhibit low interparticle distances.
Therefore, the observed features indicate the presence of a hierarchical organization in the energy levels, contingent upon
$\Phi_0^i$. In this context, the lowest energy can be attributed
to the outermost EC, which occurs in
the flat tail of the potential energy surface. Therefore, as will be demonstrated in the subsequent section, a certain degree 
of spatial and energetic isolation of the lowest energy state occurs.
This observation provides a plausible explanation for the discernible jumps in the increase
of the lowest energy upon transitioning to higher $N$ (for $N$ of 3-4, 5-6 and 7-8),
corresponding to the aforementioned vanishing of the highest clusters of $\Phi_0^i$ (see Fig. \ref{fig:cluster_plus_histogram}(a-g)).
Moreover, for each $N \geq 5$, a small cluster of low energy
levels, containing the lowest energy, is evident, exhibiting a gap to the other levels.
This can be attributed to the presence of ECs with large $\Phi_0^i$, that
are energetically somewhat isolated. 

\subsection{Correlations between spatial and energetic properties}
\begin{figure}[t]
    \centering
    \includegraphics[width=\columnwidth]{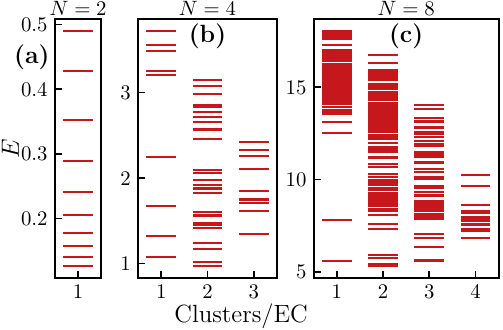}
    \caption{The energy levels of the ECs for \textbf{(a)} $N=2$, \textbf{(b)} $N=4$ and \textbf{(c)} $N=8$,
             contingent on the above employed metric of Clusters/EC (see Fig. \ref{fig:num_clusters_ec}(a-h)).}
    \label{fig:degeneracy}
\end{figure}
\begin{figure*}[t]
    \centering
    \includegraphics[width=\textwidth]{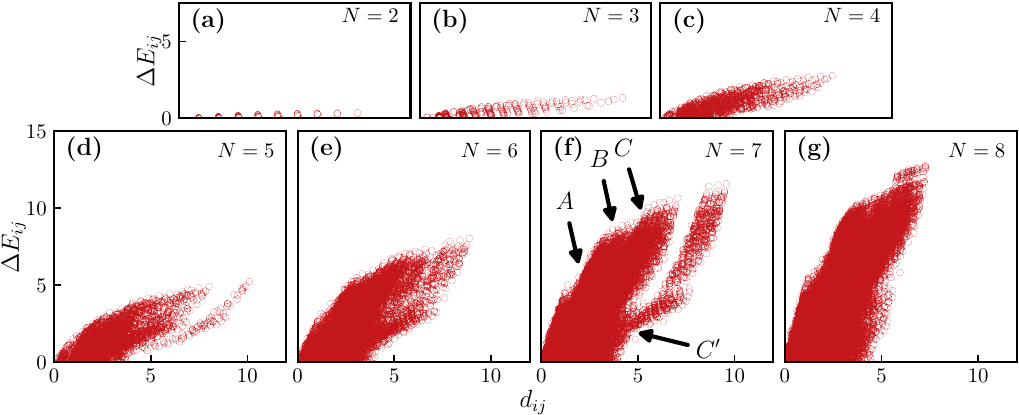}
    \caption{\textbf{(a-g)} The energy differences $\Delta E_{ij} = \lvert V(\mathbf{\Phi}_0^i) - V(\mathbf{\Phi}_0^j)\rvert$
             contingent on the pairwise distances $d_{ij}$ for $i\neq j$ and $N$ in the range 2-8.}
    \label{fig:distance_ij_energy_corr}
\end{figure*}
As indicated above, a correlation exists between the energy
levels and the spatial properties of the ECs. This correlation will be analyzed in this
subsection. Fig. \ref{fig:degeneracy}(a-c) illustrates the energy levels of the ECs for $N=2,4,8$, contingent on the
above employed metric of Clusters/EC. In the case of $N=4,8$, a hierarchical organization of energy levels
emerges, characterized by a pronounced decrease in maximum energies with an increase in Clusters/EC.
Concomitantly, the value of the minimum energy level overall increases along this crossover after reaching
the energy of the lowest energy state for Clusters/EC equal to 2, thereby leading to a decrease in the energy range $E_{\mathrm{max}} - E_{\mathrm{min}}$ (see Fig. \ref{fig:degeneracy}(b,c)).

In the event that the number of clusters is equal to 1, a transition from a regular distribution for $N=2$ to an irregular distribution with the formation
of two separated energy clusters for low and for high energy levels for $N=4,8$ is discernible.
In accordance with Fig. \ref{fig:num_clusters_ec}(b,d,h), the energies within the upper energy cluster are
assignable to ECs comprised by interparticle distancs from the first cluster of $\Phi_0^i$ while
the ECs whose energies are in the lower energy cluster are constituted by interparticle distances
from one of the other occupiable cluster. The developing gap is further increasing with $N$. Additionally,
as the total number of interparticle distances situated in the first cluster increases by $N$,
the upper energy cluster widens and the lower one narrows.

In a similar manner, the energy distributions for Clusters/EC equal to 2 and 3 demonstrate an increasing
level density for higher energies. This is associated with the increasing probability that the interparticle
distances of these ECs predominantly are in lower clusters (see Fig. \ref{fig:num_clusters_ec}(b,d,h)).
For the Clusters/EC equal to 4, the wide distribution of $C(\Phi_0^i)$ necessitates that the ECs possess
interparticle distances corresponding to higher clusters, thus showing an elevated level
density for low energies. The same argumentation can be applied to explain the increase
in the minimum energy level by the number of Clusters/EC. 
The ECs must also contain small interparticle distances due to the limited number
of clusters in general, which precludes markedly low energies. Nevertheless, the ground
state energy emerges for Clusters/EC equal to 2 because ECs 
with merely interparticle distances from the highest occupiable cluster do not occur.
Ultimately, we can conclude 
(in conjunction with Fig. \ref{fig:cluster_correlation}(a-f) and Fig. \ref{fig:pos_i_cluster}(a-f)),
that for the ground state $\Phi_0^1$ and $\Phi_0^{N-1}$ are in the highest possible cluster, and
$\Phi_2,\ldots,\Phi_{N-2}$ in the neighboring lower cluster.

The energies of the ECs are strongly correlated
to the interparticle distances, thereby establishing an energetic hierarchy. 
However, the stringency of this hierarchy diminishes with an increase in $N$.
For instance, for $N=8$, ECs comprising interparticle distances which are distributed
across two or three clusters can attain energies comparable of those ECs consisting solely
of interparticle distances within the first cluster. This is counterintuitive, but it
should be noted that the large width of the first cluster allows for a high variety of
small interparticle distances, leading for example to $\mathbf{\Phi}_0^1 \approx (1.53, 0.23, 0.64, 0.52, 0.19, 0.19, 0.39)$
and $\mathbf{\Phi}_0^2 \approx (0.79, 0.41, 0.41, 0.83, 0.79, 0.75, 0.71)$, with 
$E(\mathbf{\Phi}_0^1) > E(\mathbf{\Phi}_0^2)$.

In order to expand the correlational analysis to the $(N-1)$-dimensional potential landscape,
Fig. \ref{fig:distance_ij_energy_corr}(a-g) shows the energy difference
$\Delta E_{ij} = \lvert V(\mathbf{\Phi}_0^i) - V(\mathbf{\Phi}_0^j)\rvert$ in dependence on the
pairwise distances $d_{ij}$ (see Fig. \ref{fig:distances_configuration_space}(a)).
A correlation is manifested in the manner, that
ECs that are in a large spatial distance also rather show a great difference in their energies,
thereby underpinning the established energy hierarchy.
In accordance with Fig. \ref{fig:energy_distribution}(a) and Fig. \ref{fig:distances_configuration_space}(a), the
maximum energy level exhibits an upward trend, while the maximum pairwise distance demonstrates a downward trend
as $N$ increases, 
resulting in a more and more pronounced increase in $\Delta E_{ij}$ by $d_{ij}$ for the
crossover from $N=2$ to $N=8$.

For $N \geq 5$, the maximum value of $\Delta E_{ij}$ (i.e. the upper envelope of the correlation plot)
exhibit a pronounced increase for small
values of $d_{ij}$ (exemplarily marked as region $A$ in Fig. \ref{fig:distance_ij_energy_corr}(f)).
This increase subsequently flattens out for spatially more separated ECs ($B$). 
The high values of $\Delta E_{ij}$ can be assigned to ECs where one of them has
a high corresponding energy and
interparticle distances that are primarily in the first cluster. Thus, the flattening out of
$\Delta E_{ij}$ implies a more level potential energy surface for intermediate to large interparticle
distances and a highly oscillatory surface for small interparticle distances. 
However, in the flattened
region discontinuous increases in the maximum value of $\Delta E_{ij}$ are discernible, particularly
at $d_{ij} \approx 5$ for $N \geq 7$ ($C$). At this particular value of $d_{ij}$, it is intriguing
that also for the minimum value of $\Delta E_{ij}$ a discontinuity occurs $(C^\prime)$, and
a separated branch of $\Delta E_{ij}$ values is formed. This behavior is also insinuated
for $N=5,6$. For $N=8$ the separated branch is not visible anymore. However, the results
for $N=8$ should be treated with caution due to possibly missing ECs (see section \ref{chap:three}).
Fig. \ref{fig:energy_distribution}(a) demonstrates the occurence of minor, isolated,
low-energy clusters for $N\geq 5$, corresponding to ECs with large interparticle
distances (including the outermost EC).
The formation of the separate branches in Fig. \ref{fig:distance_ij_energy_corr}(d-f)
can be attributed to the Euclidean distances and energy differences between these ECs
and the EC corresponding to the uppermost energy level.
The formation of the branch is indicative of both energetic and spatial isolation of these ECs.
Furthermore, within the context of an approximately continuous range of values $d_{ij}$
(in our resolution), the occuring discontinuities in the maximum and minimum
value of $\Delta E_{ij}$ indicate a highly non-linear oscillatory behavior
of the potential landscape which increases by $N$.

A strong correlation is exhibited between the spatial and energetic properties, whereby
the energies of the ECs show an overall hierarchical ordering contingent on the
interparticle distances, respectively its location within the configuration space.
As $N$ increases the hierarchy is subject to more and more fluctuations, especially
for ECs with small $\Phi_0^i$, for which significant energetic differences can exist,
despite their spatial proximity. The correlations demonstrate that the potential
landscape is strongly oscillatory for small $\Phi_0^i$ but flattens
out for an increase in their coordinates, resulting in spatially more
isolated ECs.

\section{Summary and Conclusions}\label{chap:five}
Charged particles in a helical confinement have previously been demonstrated
to exhibit a variety of intriguing properties which are highly tunable 
according to the helical parameters \cite{kibisElectronelectronInteractionSpiral1992,
schmelcherEffectiveLongrangeInteractions2011,pedersenFormationClassicalCrystals2014,pedersenQuantumFewbodyBound2016,
siemensTunableOrderHelically2020,plettenbergLocalEquilibriaState2017,
gloyDrivenToroidalHelix2022,zampetakiClassicalScatteringCharged2013,
siemensClassicalScatteringFragmentation2025,zampetakiDynamicsNonlinearExcitations2015,
zampetakiDegeneracyInversionBand2015}.
In this work, we addressed a gap in the literature by providing
a statistical and correlational analysis of the ECs of a few-body system
(up to eight particles) on a straight, infinite helix.
Evidence has demonstrated a power-law relationship between the number
of ECs and the number of particles $(N^{2.7})$.
According to the repulsion within a helical winding,
the observed values of interparticle distances in ECs, clusterize around values
of approximately odd multiples of half a helical winding.
Correlations of the interparticle distance among each other within an EC,
have revealed that neighboring interparticle distances are likely to be
in the same or in a neighboring cluster and that a high probability for the sequences of ECs
to be ordered hierarchically or symmetrically in $\Phi_0^i$ and $\Phi_0^{N-i}$ exists.
The clustering in conjunction with the analysis of pairwise distances
of ECs within the high-dimensional potential landscape has yielded
a crossover from regularly distributed ECs in the two-body
case to a heightened degree of disorder, concomitant with the increase
of $N$ and an increase in ECs comprised by small interparticle distances,
thus indicating the presence of a strongly oscillatory
potential energy surface.
Furthermore, the strict hierarchical ordering of energies corresponding
to ECs, is subject to increasing fluctuations
as $N$ increases, even exhibiting significant energy gaps between ECs
in close spatial proximity.

A natural subsequent step would be to examine the dynamical properties
of the helical few-body chain through a
harmonic approximation. From a statistical viewpoint an extensive analysis 
of the phase space would be intriguing.
Moreover, this could be combined with a more sophisticated optimization
approach, utilizing the accumulated empiricism to study
the many-body system.
In fact, incomplete simulations, extending to $N=20$, show the persistence of many of the 
features previously outlined, including the clustering of interparticle distances, the diminution 
of the width of higher clusters and the concomitant augmentation of the first cluster, which 
consequently predominates in the statistical and correlational analysis. Additional persistent 
features are for instance a high degree of irregularity within the spatial distribution of ECs as 
well as increasing energies. Nonetheless, it would be unwarranted to expect that the features of 
the system presented here would be equally present in systems with a significantly larger number 
of particles, which therefore could give rise to newly emerging features.

In terms of perspective, the development of the quantum physics of the
helical many-body chain represents an intriguing extension, concerning
the structure of eigenstates and potential novel phenomena for the
dynamical behavior.
First investigations on quantum few-body bound states
in helical confinement have been already done in Ref. \cite{pedersenQuantumFewbodyBound2016}.
Additionally, the vast number of metastable
ECs that emerge due to the "competition" of curvature and repulsion
bear a resemblance to the frustrated interactions in spin glasses
and indicates a similar behavior for the helical many-body chain.

Ultimately, experimental realizations for similar setups were demonstrated
to be feasible. For instance, a double helical dipole trap for cold
neutral atoms, enabling the experimental investigation of the properties
of a helical many-body chain in the domain of cold atom physics \cite{reitzNanofiberbasedDoublehelixDipole2012}.

\begin{acknowledgments}
J.M.D. and P.S. appreciate very much helpful discussions
with Ansgar Siemens in the early stages of this project.
\end{acknowledgments}


\begin{thebibliography}{40}%
\makeatletter
\providecommand \@ifxundefined [1]{%
 \@ifx{#1\undefined}
}%
\providecommand \@ifnum [1]{%
 \ifnum #1\expandafter \@firstoftwo
 \else \expandafter \@secondoftwo
 \fi
}%
\providecommand \@ifx [1]{%
 \ifx #1\expandafter \@firstoftwo
 \else \expandafter \@secondoftwo
 \fi
}%
\providecommand \natexlab [1]{#1}%
\providecommand \enquote  [1]{``#1''}%
\providecommand \bibnamefont  [1]{#1}%
\providecommand \bibfnamefont [1]{#1}%
\providecommand \citenamefont [1]{#1}%
\providecommand \href@noop [0]{\@secondoftwo}%
\providecommand \href [0]{\begingroup \@sanitize@url \@href}%
\providecommand \@href[1]{\@@startlink{#1}\@@href}%
\providecommand \@@href[1]{\endgroup#1\@@endlink}%
\providecommand \@sanitize@url [0]{\catcode `\\12\catcode `\$12\catcode
  `\&12\catcode `\#12\catcode `\^12\catcode `\_12\catcode `\%12\relax}%
\providecommand \@@startlink[1]{}%
\providecommand \@@endlink[0]{}%
\providecommand \url  [0]{\begingroup\@sanitize@url \@url }%
\providecommand \@url [1]{\endgroup\@href {#1}{\urlprefix }}%
\providecommand \urlprefix  [0]{URL }%
\providecommand \Eprint [0]{\href }%
\providecommand \doibase [0]{https://doi.org/}%
\providecommand \selectlanguage [0]{\@gobble}%
\providecommand \bibinfo  [0]{\@secondoftwo}%
\providecommand \bibfield  [0]{\@secondoftwo}%
\providecommand \translation [1]{[#1]}%
\providecommand \BibitemOpen [0]{}%
\providecommand \bibitemStop [0]{}%
\providecommand \bibitemNoStop [0]{.\EOS\space}%
\providecommand \EOS [0]{\spacefactor3000\relax}%
\providecommand \BibitemShut  [1]{\csname bibitem#1\endcsname}%
\let\auto@bib@innerbib\@empty
\bibitem [{\citenamefont {Watson}\ and\ \citenamefont
  {Crick}(1953)}]{watsonMolecularStructureNucleic1953}%
  \BibitemOpen
  \bibfield  {author} {\bibinfo {author} {\bibfnamefont {J.~D.}\ \bibnamefont
  {Watson}}\ and\ \bibinfo {author} {\bibfnamefont {F.~H.~C.}\ \bibnamefont
  {Crick}},\ }\bibfield  {title} {\bibinfo {title} {Molecular {{Structure}} of
  {{Nucleic Acids}}: {{A Structure}} for {{Deoxyribose Nucleic Acid}}},\ }\href
  {https://www.nature.com/articles/171737a0} {\bibfield  {journal} {\bibinfo
  {journal} {Nature}\ }\textbf {\bibinfo {volume} {171}},\ \bibinfo {pages}
  {737} (\bibinfo {year} {1953})}\BibitemShut {NoStop}%
\bibitem [{\citenamefont {Travers}\ and\ \citenamefont
  {Muskhelishvili}(2015)}]{traversDNAStructureFunction2015}%
  \BibitemOpen
  \bibfield  {author} {\bibinfo {author} {\bibfnamefont {A.}~\bibnamefont
  {Travers}}\ and\ \bibinfo {author} {\bibfnamefont {G.}~\bibnamefont
  {Muskhelishvili}},\ }\bibfield  {title} {\bibinfo {title} {{{DNA}} structure
  and function},\ }\href {https://doi.org/10.1111/febs.13307} {\bibfield
  {journal} {\bibinfo  {journal} {FEBS J.}\ }\textbf {\bibinfo {volume}
  {282}},\ \bibinfo {pages} {2279} (\bibinfo {year} {2015})}\BibitemShut
  {NoStop}%
\bibitem [{\citenamefont {Pauling}\ \emph {et~al.}(1951)\citenamefont
  {Pauling}, \citenamefont {Corey},\ and\ \citenamefont
  {Branson}}]{paulingStructureProteinsTwo1951}%
  \BibitemOpen
  \bibfield  {author} {\bibinfo {author} {\bibfnamefont {L.}~\bibnamefont
  {Pauling}}, \bibinfo {author} {\bibfnamefont {R.~B.}\ \bibnamefont {Corey}},\
  and\ \bibinfo {author} {\bibfnamefont {H.~R.}\ \bibnamefont {Branson}},\
  }\bibfield  {title} {\bibinfo {title} {The structure of proteins; two
  hydrogen-bonded helical configurations of the polypeptide chain},\ }\href
  {https://doi.org/10.1073/pnas.37.4.205} {\bibfield  {journal} {\bibinfo
  {journal} {Proc. Natl. Acad. Sci. U.S.A}\ }\textbf {\bibinfo {volume} {37}},\
  \bibinfo {pages} {205} (\bibinfo {year} {1951})}\BibitemShut {NoStop}%
\bibitem [{\citenamefont {Pauling}\ and\ \citenamefont
  {Corey}(1951)}]{paulingConfigurationsPolypeptideChains1951}%
  \BibitemOpen
  \bibfield  {author} {\bibinfo {author} {\bibfnamefont {L.}~\bibnamefont
  {Pauling}}\ and\ \bibinfo {author} {\bibfnamefont {R.~B.}\ \bibnamefont
  {Corey}},\ }\bibfield  {title} {\bibinfo {title} {Configurations of
  {{Polypeptide Chains With Favored Orientations Around Single Bonds}}},\
  }\href {https://www.ncbi.nlm.nih.gov/pmc/articles/PMC1063460/} {\bibfield
  {journal} {\bibinfo  {journal} {Proc. Natl. Acad. Sci. U.S.A}\ }\textbf
  {\bibinfo {volume} {37}},\ \bibinfo {pages} {729} (\bibinfo {year}
  {1951})}\BibitemShut {NoStop}%
\bibitem [{\citenamefont {Chothia}\ \emph {et~al.}(1977)\citenamefont
  {Chothia}, \citenamefont {Levitt},\ and\ \citenamefont
  {Richardson}}]{chothiaStructureProteinsPacking1977}%
  \BibitemOpen
  \bibfield  {author} {\bibinfo {author} {\bibfnamefont {C.}~\bibnamefont
  {Chothia}}, \bibinfo {author} {\bibfnamefont {M.}~\bibnamefont {Levitt}},\
  and\ \bibinfo {author} {\bibfnamefont {D.}~\bibnamefont {Richardson}},\
  }\bibfield  {title} {\bibinfo {title} {Structure of proteins: Packing of
  alpha-helices and pleated sheets.},\ }\href
  {https://www.pnas.org/doi/abs/10.1073/pnas.74.10.4130} {\bibfield  {journal}
  {\bibinfo  {journal} {Proc. Natl. Acad. Sci. U.S.A.}\ }\textbf {\bibinfo
  {volume} {74}},\ \bibinfo {pages} {4130} (\bibinfo {year}
  {1977})}\BibitemShut {NoStop}%
\bibitem [{\citenamefont {Cohen}\ and\ \citenamefont
  {Parry}(1990)}]{cohenAlphahelicalCoiledCoils1990}%
  \BibitemOpen
  \bibfield  {author} {\bibinfo {author} {\bibfnamefont {C.}~\bibnamefont
  {Cohen}}\ and\ \bibinfo {author} {\bibfnamefont {D.~A.}\ \bibnamefont
  {Parry}},\ }\bibfield  {title} {\bibinfo {title} {Alpha-helical coiled coils
  and bundles: How to design an alpha-helical protein},\ }\href
  {https://doi.org/10.1002/prot.340070102} {\bibfield  {journal} {\bibinfo
  {journal} {Proteins}\ }\textbf {\bibinfo {volume} {7}},\ \bibinfo {pages} {1}
  (\bibinfo {year} {1990})}\BibitemShut {NoStop}%
\bibitem [{\citenamefont {Gutierrez}\ \emph {et~al.}(2012)\citenamefont
  {Gutierrez}, \citenamefont {D{\'i}az}, \citenamefont {Naaman},\ and\
  \citenamefont {Cuniberti}}]{gutierrezSpinselectiveTransportHelical2012}%
  \BibitemOpen
  \bibfield  {author} {\bibinfo {author} {\bibfnamefont {R.}~\bibnamefont
  {Gutierrez}}, \bibinfo {author} {\bibfnamefont {E.}~\bibnamefont {D{\'i}az}},
  \bibinfo {author} {\bibfnamefont {R.}~\bibnamefont {Naaman}},\ and\ \bibinfo
  {author} {\bibfnamefont {G.}~\bibnamefont {Cuniberti}},\ }\bibfield  {title}
  {\bibinfo {title} {Spin-selective transport through helical molecular
  systems},\ }\href {https://link.aps.org/doi/10.1103/PhysRevB.85.081404}
  {\bibfield  {journal} {\bibinfo  {journal} {Phys. Rev. B}\ }\textbf {\bibinfo
  {volume} {85}},\ \bibinfo {pages} {081404} (\bibinfo {year}
  {2012})}\BibitemShut {NoStop}%
\bibitem [{\citenamefont {Guo}\ and\ \citenamefont
  {Sun}(2012)}]{guoSpinSelectiveTransportElectrons2012}%
  \BibitemOpen
  \bibfield  {author} {\bibinfo {author} {\bibfnamefont {A.-M.}\ \bibnamefont
  {Guo}}\ and\ \bibinfo {author} {\bibfnamefont {Q.-F.}\ \bibnamefont {Sun}},\
  }\bibfield  {title} {\bibinfo {title} {Spin-{{Selective Transport}} of
  {{Electrons}} in {{DNA Double Helix}}},\ }\href
  {https://link.aps.org/doi/10.1103/PhysRevLett.108.218102} {\bibfield
  {journal} {\bibinfo  {journal} {Phys. Rev. Lett.}\ }\textbf {\bibinfo
  {volume} {108}},\ \bibinfo {pages} {218102} (\bibinfo {year}
  {2012})}\BibitemShut {NoStop}%
\bibitem [{\citenamefont {Guo}\ and\ \citenamefont
  {Sun}(2014)}]{guoSpindependentElectronTransport2014}%
  \BibitemOpen
  \bibfield  {author} {\bibinfo {author} {\bibfnamefont {A.-M.}\ \bibnamefont
  {Guo}}\ and\ \bibinfo {author} {\bibfnamefont {Q.-F.}\ \bibnamefont {Sun}},\
  }\bibfield  {title} {\bibinfo {title} {Spin-dependent electron transport in
  protein-like single-helical molecules},\ }\href
  {https://www.pnas.org/doi/full/10.1073/pnas.1407716111} {\bibfield  {journal}
  {\bibinfo  {journal} {Proc. Natl. Acad. Sci. U.S.A.}\ }\textbf {\bibinfo
  {volume} {111}},\ \bibinfo {pages} {11658} (\bibinfo {year}
  {2014})}\BibitemShut {NoStop}%
\bibitem [{\citenamefont {Yoo}\ \emph {et~al.}(2001)\citenamefont {Yoo},
  \citenamefont {Ha}, \citenamefont {Lee}, \citenamefont {Park}, \citenamefont
  {Kim}, \citenamefont {Kim}, \citenamefont {Lee}, \citenamefont {Kawai},\ and\
  \citenamefont {Choi}}]{yooElectricalConductionPolydAPolydT2001}%
  \BibitemOpen
  \bibfield  {author} {\bibinfo {author} {\bibfnamefont {K.-H.}\ \bibnamefont
  {Yoo}}, \bibinfo {author} {\bibfnamefont {D.~H.}\ \bibnamefont {Ha}},
  \bibinfo {author} {\bibfnamefont {J.-O.}\ \bibnamefont {Lee}}, \bibinfo
  {author} {\bibfnamefont {J.~W.}\ \bibnamefont {Park}}, \bibinfo {author}
  {\bibfnamefont {J.}~\bibnamefont {Kim}}, \bibinfo {author} {\bibfnamefont
  {J.~J.}\ \bibnamefont {Kim}}, \bibinfo {author} {\bibfnamefont {H.-Y.}\
  \bibnamefont {Lee}}, \bibinfo {author} {\bibfnamefont {T.}~\bibnamefont
  {Kawai}},\ and\ \bibinfo {author} {\bibfnamefont {H.~Y.}\ \bibnamefont
  {Choi}},\ }\bibfield  {title} {\bibinfo {title} {Electrical {{Conduction}}
  through {{Poly}}({{dA}})-{{Poly}}({{dT}}) and
  {{Poly}}({{dG}})-{{Poly}}({{dC}}) {{DNA Molecules}}},\ }\href
  {https://link.aps.org/doi/10.1103/PhysRevLett.87.198102} {\bibfield
  {journal} {\bibinfo  {journal} {Phys. Rev. Lett.}\ }\textbf {\bibinfo
  {volume} {87}},\ \bibinfo {pages} {198102} (\bibinfo {year}
  {2001})}\BibitemShut {NoStop}%
\bibitem [{\citenamefont {Malyshev}(2007)}]{malyshevDNADoubleHelices2007}%
  \BibitemOpen
  \bibfield  {author} {\bibinfo {author} {\bibfnamefont {A.~V.}\ \bibnamefont
  {Malyshev}},\ }\bibfield  {title} {\bibinfo {title} {{{DNA Double Helices}}
  for {{Single Molecule Electronics}}},\ }\href
  {https://link.aps.org/doi/10.1103/PhysRevLett.98.096801} {\bibfield
  {journal} {\bibinfo  {journal} {Phys. Rev. Lett.}\ }\textbf {\bibinfo
  {volume} {98}},\ \bibinfo {pages} {096801} (\bibinfo {year}
  {2007})}\BibitemShut {NoStop}%
\bibitem [{\citenamefont {Ren}\ and\ \citenamefont
  {Gao}(2014)}]{renReviewHelicalNanostructures2014}%
  \BibitemOpen
  \bibfield  {author} {\bibinfo {author} {\bibfnamefont {Z.}~\bibnamefont
  {Ren}}\ and\ \bibinfo {author} {\bibfnamefont {P.-X.}\ \bibnamefont {Gao}},\
  }\bibfield  {title} {\bibinfo {title} {A review of helical nanostructures:
  Growth theories, synthesis strategies and properties},\ }\href
  {https://pubs.rsc.org/en/content/articlelanding/2014/nr/c4nr00330f}
  {\bibfield  {journal} {\bibinfo  {journal} {Nanoscale}\ }\textbf {\bibinfo
  {volume} {6}},\ \bibinfo {pages} {9366} (\bibinfo {year} {2014})}\BibitemShut
  {NoStop}%
\bibitem [{\citenamefont {Huang}\ and\ \citenamefont
  {Mei}(2015)}]{huangHelicesMicroworldMaterials2015}%
  \BibitemOpen
  \bibfield  {author} {\bibinfo {author} {\bibfnamefont {G.}~\bibnamefont
  {Huang}}\ and\ \bibinfo {author} {\bibfnamefont {Y.}~\bibnamefont {Mei}},\
  }\bibfield  {title} {\bibinfo {title} {Helices in micro-world: {{Materials}},
  properties, and applications},\ }\href
  {https://www.sciencedirect.com/science/article/pii/S2352847815000726}
  {\bibfield  {journal} {\bibinfo  {journal} {J. Mater.}\ }\textbf {\bibinfo
  {volume} {1}},\ \bibinfo {pages} {296} (\bibinfo {year} {2015})}\BibitemShut
  {NoStop}%
\bibitem [{\citenamefont {Lau}\ \emph {et~al.}(2006)\citenamefont {Lau},
  \citenamefont {Lu},\ and\ \citenamefont
  {Hui}}]{lauCoiledCarbonNanotubes2006}%
  \BibitemOpen
  \bibfield  {author} {\bibinfo {author} {\bibfnamefont {K.~T.}\ \bibnamefont
  {Lau}}, \bibinfo {author} {\bibfnamefont {M.}~\bibnamefont {Lu}},\ and\
  \bibinfo {author} {\bibfnamefont {D.}~\bibnamefont {Hui}},\ }\bibfield
  {title} {\bibinfo {title} {Coiled carbon nanotubes: {{Synthesis}} and their
  potential applications in advanced composite structures},\ }\href
  {https://www.sciencedirect.com/science/article/pii/S1359836806000187}
  {\bibfield  {journal} {\bibinfo  {journal} {Comp. B.: Eng.}\ }\textbf
  {\bibinfo {volume} {37}},\ \bibinfo {pages} {437} (\bibinfo {year}
  {2006})}\BibitemShut {NoStop}%
\bibitem [{\citenamefont {Shaikjee}\ and\ \citenamefont
  {Coville}(2012)}]{shaikjeeSynthesisPropertiesUses2012}%
  \BibitemOpen
  \bibfield  {author} {\bibinfo {author} {\bibfnamefont {A.}~\bibnamefont
  {Shaikjee}}\ and\ \bibinfo {author} {\bibfnamefont {N.~J.}\ \bibnamefont
  {Coville}},\ }\bibfield  {title} {\bibinfo {title} {The synthesis, properties
  and uses of carbon materials with helical morphology},\ }\href
  {https://www.sciencedirect.com/science/article/pii/S2090123211000725}
  {\bibfield  {journal} {\bibinfo  {journal} {J. Adv. Res.}\ }\textbf {\bibinfo
  {volume} {3}},\ \bibinfo {pages} {195} (\bibinfo {year} {2012})}\BibitemShut
  {NoStop}%
\bibitem [{\citenamefont {Gibbs}\ \emph {et~al.}(2014)\citenamefont {Gibbs},
  \citenamefont {Mark}, \citenamefont {Lee}, \citenamefont {Eslami},
  \citenamefont {Schamel},\ and\ \citenamefont
  {Fischer}}]{gibbsNanohelicesShadowGrowth2014}%
  \BibitemOpen
  \bibfield  {author} {\bibinfo {author} {\bibfnamefont {J.~G.}\ \bibnamefont
  {Gibbs}}, \bibinfo {author} {\bibfnamefont {A.~G.}\ \bibnamefont {Mark}},
  \bibinfo {author} {\bibfnamefont {T.-C.}\ \bibnamefont {Lee}}, \bibinfo
  {author} {\bibfnamefont {S.}~\bibnamefont {Eslami}}, \bibinfo {author}
  {\bibfnamefont {D.}~\bibnamefont {Schamel}},\ and\ \bibinfo {author}
  {\bibfnamefont {P.}~\bibnamefont {Fischer}},\ }\bibfield  {title} {\bibinfo
  {title} {Nanohelices by shadow growth},\ }\href
  {https://pubs.rsc.org/en/content/articlelanding/2014/nr/c4nr00403e}
  {\bibfield  {journal} {\bibinfo  {journal} {Nanoscale}\ }\textbf {\bibinfo
  {volume} {6}},\ \bibinfo {pages} {9457} (\bibinfo {year} {2014})}\BibitemShut
  {NoStop}%
\bibitem [{\citenamefont {Liu}\ \emph {et~al.}(2014)\citenamefont {Liu},
  \citenamefont {Shen}, \citenamefont {Wang}, \citenamefont {Kuzyk},\ and\
  \citenamefont {Ding}}]{liuHelicalNanostructuresBased2014}%
  \BibitemOpen
  \bibfield  {author} {\bibinfo {author} {\bibfnamefont {H.}~\bibnamefont
  {Liu}}, \bibinfo {author} {\bibfnamefont {X.}~\bibnamefont {Shen}}, \bibinfo
  {author} {\bibfnamefont {Z.-G.}\ \bibnamefont {Wang}}, \bibinfo {author}
  {\bibfnamefont {A.}~\bibnamefont {Kuzyk}},\ and\ \bibinfo {author}
  {\bibfnamefont {B.}~\bibnamefont {Ding}},\ }\bibfield  {title} {\bibinfo
  {title} {Helical nanostructures based on {{DNA}} self-assembly},\ }\href
  {https://pubs.rsc.org/en/content/articlelanding/2014/nr/c3nr06913c}
  {\bibfield  {journal} {\bibinfo  {journal} {Nanoscale}\ }\textbf {\bibinfo
  {volume} {6}},\ \bibinfo {pages} {9331} (\bibinfo {year} {2014})}\BibitemShut
  {NoStop}%
\bibitem [{\citenamefont {Kohlstedt}\ \emph {et~al.}(2007)\citenamefont
  {Kohlstedt}, \citenamefont {Solis}, \citenamefont {Vernizzi},\ and\
  \citenamefont {De~La~Cruz}}]{kohlstedtSpontaneousChiralityLongrange2007}%
  \BibitemOpen
  \bibfield  {author} {\bibinfo {author} {\bibfnamefont {K.~L.}\ \bibnamefont
  {Kohlstedt}}, \bibinfo {author} {\bibfnamefont {F.~J.}\ \bibnamefont
  {Solis}}, \bibinfo {author} {\bibfnamefont {G.}~\bibnamefont {Vernizzi}},\
  and\ \bibinfo {author} {\bibfnamefont {M.~O.}\ \bibnamefont {De~La~Cruz}},\
  }\bibfield  {title} {\bibinfo {title} {Spontaneous chirality via long-range
  electrostatic forces},\ }\href
  {http://www.scopus.com/inward/record.url?scp=34547214389&partnerID=8YFLogxK}
  {\bibfield  {journal} {\bibinfo  {journal} {Phys. Rev. Lett.}\ }\textbf
  {\bibinfo {volume} {99}} (\bibinfo {year} {2007})}\BibitemShut {NoStop}%
\bibitem [{\citenamefont {Vernizzi}\ \emph {et~al.}(2009)\citenamefont
  {Vernizzi}, \citenamefont {Kohlstedt},\ and\ \citenamefont {de~la
  Cruz}}]{vernizziElectrostaticOriginChiral2009}%
  \BibitemOpen
  \bibfield  {author} {\bibinfo {author} {\bibfnamefont {G.}~\bibnamefont
  {Vernizzi}}, \bibinfo {author} {\bibfnamefont {K.~L.}\ \bibnamefont
  {Kohlstedt}},\ and\ \bibinfo {author} {\bibfnamefont {M.~O.}\ \bibnamefont
  {de~la Cruz}},\ }\bibfield  {title} {\bibinfo {title} {The electrostatic
  origin of chiral patterns on nanofibers},\ }\href
  {https://pubs.rsc.org/en/content/articlelanding/2009/sm/b814583k} {\bibfield
  {journal} {\bibinfo  {journal} {Soft Matter}\ }\textbf {\bibinfo {volume}
  {5}},\ \bibinfo {pages} {736} (\bibinfo {year} {2009})}\BibitemShut {NoStop}%
\bibitem [{\citenamefont {Srebnik}\ and\ \citenamefont
  {Douglas}(2011)}]{srebnikSelfassemblyChargedParticles2011}%
  \BibitemOpen
  \bibfield  {author} {\bibinfo {author} {\bibfnamefont {S.}~\bibnamefont
  {Srebnik}}\ and\ \bibinfo {author} {\bibfnamefont {J.~F.}\ \bibnamefont
  {Douglas}},\ }\bibfield  {title} {\bibinfo {title} {Self-assembly of charged
  particles on nanotubes and the emergence of particle rings, chains, ribbons
  and chiral sheets},\ }\href
  {https://pubs.rsc.org/en/content/articlelanding/2011/sm/c1sm05168g}
  {\bibfield  {journal} {\bibinfo  {journal} {Soft Matter}\ }\textbf {\bibinfo
  {volume} {7}},\ \bibinfo {pages} {6897} (\bibinfo {year} {2011})}\BibitemShut
  {NoStop}%
\bibitem [{\citenamefont {Gao}\ \emph {et~al.}(2006)\citenamefont {Gao},
  \citenamefont {Mai},\ and\ \citenamefont
  {Wang}}]{gaoSuperelasticityNanofractureMechanics2006}%
  \BibitemOpen
  \bibfield  {author} {\bibinfo {author} {\bibfnamefont {P.~X.}\ \bibnamefont
  {Gao}}, \bibinfo {author} {\bibfnamefont {W.}~\bibnamefont {Mai}},\ and\
  \bibinfo {author} {\bibfnamefont {Z.~L.}\ \bibnamefont {Wang}},\ }\bibfield
  {title} {\bibinfo {title} {Superelasticity and {{Nanofracture Mechanics}} of
  {{ZnO Nanohelices}}},\ }\href {https://doi.org/10.1021/nl061943i} {\bibfield
  {journal} {\bibinfo  {journal} {Nano Lett.}\ }\textbf {\bibinfo {volume}
  {6}},\ \bibinfo {pages} {2536} (\bibinfo {year} {2006})}\BibitemShut
  {NoStop}%
\bibitem [{\citenamefont {Tasco}\ \emph {et~al.}(2016)\citenamefont {Tasco},
  \citenamefont {Esposito}, \citenamefont {Todisco}, \citenamefont {Benedetti},
  \citenamefont {Cuscun{\`a}}, \citenamefont {Sanvitto},\ and\ \citenamefont
  {Passaseo}}]{tascoThreedimensionalNanohelicesChiral2016}%
  \BibitemOpen
  \bibfield  {author} {\bibinfo {author} {\bibfnamefont {V.}~\bibnamefont
  {Tasco}}, \bibinfo {author} {\bibfnamefont {M.}~\bibnamefont {Esposito}},
  \bibinfo {author} {\bibfnamefont {F.}~\bibnamefont {Todisco}}, \bibinfo
  {author} {\bibfnamefont {A.}~\bibnamefont {Benedetti}}, \bibinfo {author}
  {\bibfnamefont {M.}~\bibnamefont {Cuscun{\`a}}}, \bibinfo {author}
  {\bibfnamefont {D.}~\bibnamefont {Sanvitto}},\ and\ \bibinfo {author}
  {\bibfnamefont {A.}~\bibnamefont {Passaseo}},\ }\bibfield  {title} {\bibinfo
  {title} {Three-dimensional nanohelices for chiral photonics},\ }\href
  {https://doi.org/10.1007/s00339-016-9856-6} {\bibfield  {journal} {\bibinfo
  {journal} {Appl. Phys. A}\ }\textbf {\bibinfo {volume} {122}},\ \bibinfo
  {pages} {280} (\bibinfo {year} {2016})}\BibitemShut {NoStop}%
\bibitem [{\citenamefont {Kosters}\ \emph {et~al.}(2017)\citenamefont
  {Kosters}, \citenamefont {{de Hoogh}}, \citenamefont {Zeijlemaker},
  \citenamefont {Acar}, \citenamefont {Rotenberg},\ and\ \citenamefont
  {Kuipers}}]{kostersCoreShellPlasmonic2017}%
  \BibitemOpen
  \bibfield  {author} {\bibinfo {author} {\bibfnamefont {D.}~\bibnamefont
  {Kosters}}, \bibinfo {author} {\bibfnamefont {A.}~\bibnamefont {{de Hoogh}}},
  \bibinfo {author} {\bibfnamefont {H.}~\bibnamefont {Zeijlemaker}}, \bibinfo
  {author} {\bibfnamefont {H.}~\bibnamefont {Acar}}, \bibinfo {author}
  {\bibfnamefont {N.}~\bibnamefont {Rotenberg}},\ and\ \bibinfo {author}
  {\bibfnamefont {L.}~\bibnamefont {Kuipers}},\ }\bibfield  {title} {\bibinfo
  {title} {Core--{{Shell Plasmonic Nanohelices}}},\ }\href
  {https://doi.org/10.1021/acsphotonics.7b00496} {\bibfield  {journal}
  {\bibinfo  {journal} {ACS Photonics}\ }\textbf {\bibinfo {volume} {4}},\
  \bibinfo {pages} {1858} (\bibinfo {year} {2017})}\BibitemShut {NoStop}%
\bibitem [{\citenamefont {Reitz}\ and\ \citenamefont
  {Rauschenbeutel}(2012)}]{reitzNanofiberbasedDoublehelixDipole2012}%
  \BibitemOpen
  \bibfield  {author} {\bibinfo {author} {\bibfnamefont {D.}~\bibnamefont
  {Reitz}}\ and\ \bibinfo {author} {\bibfnamefont {A.}~\bibnamefont
  {Rauschenbeutel}},\ }\bibfield  {title} {\bibinfo {title} {Nanofiber-based
  double-helix dipole trap for cold neutral atoms},\ }\href
  {https://www.sciencedirect.com/science/article/pii/S0030401812005780}
  {\bibfield  {journal} {\bibinfo  {journal} {Opt. Commun.}\ }\textbf {\bibinfo
  {volume} {285}},\ \bibinfo {pages} {4705} (\bibinfo {year}
  {2012})}\BibitemShut {NoStop}%
\bibitem [{\citenamefont {Tsytovich}\ and\ \citenamefont
  {{Gusein-zade}}(2005{\natexlab{a}})}]{tsytovichHelicalStructuresComplex2005b}%
  \BibitemOpen
  \bibfield  {author} {\bibinfo {author} {\bibfnamefont {V.~N.}\ \bibnamefont
  {Tsytovich}}\ and\ \bibinfo {author} {\bibfnamefont {N.~G.}\ \bibnamefont
  {{Gusein-zade}}},\ }\bibfield  {title} {\bibinfo {title} {Helical structures
  in complex plasma {{I}}: {{Noncollective}} interaction},\ }\href
  {https://doi.org/10.1134/1.1925789} {\bibfield  {journal} {\bibinfo
  {journal} {Plasma Phys. Rep.}\ }\textbf {\bibinfo {volume} {31}},\ \bibinfo
  {pages} {392} (\bibinfo {year} {2005}{\natexlab{a}})}\BibitemShut {NoStop}%
\bibitem [{\citenamefont {Tsytovich}\ and\ \citenamefont
  {{Gusein-zade}}(2005{\natexlab{b}})}]{tsytovichHelicalStructuresComplex2005a}%
  \BibitemOpen
  \bibfield  {author} {\bibinfo {author} {\bibfnamefont {V.~N.}\ \bibnamefont
  {Tsytovich}}\ and\ \bibinfo {author} {\bibfnamefont {N.~G.}\ \bibnamefont
  {{Gusein-zade}}},\ }\bibfield  {title} {\bibinfo {title} {Helical structures
  in complex plasma {{II}}: {{Collective}} interaction},\ }\href
  {https://doi.org/10.1134/1.2101970} {\bibfield  {journal} {\bibinfo
  {journal} {Plasma Phys. Rep.}\ }\textbf {\bibinfo {volume} {31}},\ \bibinfo
  {pages} {824} (\bibinfo {year} {2005}{\natexlab{b}})}\BibitemShut {NoStop}%
\bibitem [{\citenamefont {Kamimura}\ and\ \citenamefont
  {Ishihara}(2012)}]{kamimuraCoulombDoubleHelical2012}%
  \BibitemOpen
  \bibfield  {author} {\bibinfo {author} {\bibfnamefont {T.}~\bibnamefont
  {Kamimura}}\ and\ \bibinfo {author} {\bibfnamefont {O.}~\bibnamefont
  {Ishihara}},\ }\bibfield  {title} {\bibinfo {title} {Coulomb double helical
  structure},\ }\href {https://link.aps.org/doi/10.1103/PhysRevE.85.016406}
  {\bibfield  {journal} {\bibinfo  {journal} {Phys. Rev. E}\ }\textbf {\bibinfo
  {volume} {85}},\ \bibinfo {pages} {016406} (\bibinfo {year}
  {2012})}\BibitemShut {NoStop}%
\bibitem [{\citenamefont
  {Kibis}(1992)}]{kibisElectronelectronInteractionSpiral1992}%
  \BibitemOpen
  \bibfield  {author} {\bibinfo {author} {\bibfnamefont {O.~V.}\ \bibnamefont
  {Kibis}},\ }\bibfield  {title} {\bibinfo {title} {Electron-electron
  interaction in a spiral quantum wire},\ }\href
  {https://www.sciencedirect.com/science/article/pii/037596019290730A}
  {\bibfield  {journal} {\bibinfo  {journal} {Phys. Lett. A}\ }\textbf
  {\bibinfo {volume} {166}},\ \bibinfo {pages} {393} (\bibinfo {year}
  {1992})}\BibitemShut {NoStop}%
\bibitem [{\citenamefont
  {Schmelcher}(2011)}]{schmelcherEffectiveLongrangeInteractions2011}%
  \BibitemOpen
  \bibfield  {author} {\bibinfo {author} {\bibfnamefont {P.}~\bibnamefont
  {Schmelcher}},\ }\bibfield  {title} {\bibinfo {title} {Effective long-range
  interactions in confined curved dimensions},\ }\href
  {https://iopscience.iop.org/article/10.1209/0295-5075/95/50005} {\bibfield
  {journal} {\bibinfo  {journal} {EPL}\ }\textbf {\bibinfo {volume} {95}},\
  \bibinfo {pages} {50005} (\bibinfo {year} {2011})}\BibitemShut {NoStop}%
\bibitem [{\citenamefont {Pedersen}\ \emph {et~al.}(2014)\citenamefont
  {Pedersen}, \citenamefont {Fedorov}, \citenamefont {Jensen},\ and\
  \citenamefont {Zinner}}]{pedersenFormationClassicalCrystals2014}%
  \BibitemOpen
  \bibfield  {author} {\bibinfo {author} {\bibfnamefont {J.~K.}\ \bibnamefont
  {Pedersen}}, \bibinfo {author} {\bibfnamefont {D.~V.}\ \bibnamefont
  {Fedorov}}, \bibinfo {author} {\bibfnamefont {A.~S.}\ \bibnamefont
  {Jensen}},\ and\ \bibinfo {author} {\bibfnamefont {N.~T.}\ \bibnamefont
  {Zinner}},\ }\bibfield  {title} {\bibinfo {title} {Formation of classical
  crystals of dipolar particles in a helical geometry},\ }\href
  {https://iopscience.iop.org/article/10.1088/0953-4075/47/16/165103}
  {\bibfield  {journal} {\bibinfo  {journal} {J. Phys. B: At. Mol. Opt. Phys.}\
  }\textbf {\bibinfo {volume} {47}},\ \bibinfo {pages} {165103} (\bibinfo
  {year} {2014})}\BibitemShut {NoStop}%
\bibitem [{\citenamefont {Pedersen}\ \emph {et~al.}(2016)\citenamefont
  {Pedersen}, \citenamefont {Fedorov}, \citenamefont {Jensen},\ and\
  \citenamefont {Zinner}}]{pedersenQuantumFewbodyBound2016}%
  \BibitemOpen
  \bibfield  {author} {\bibinfo {author} {\bibfnamefont {J.~K.}\ \bibnamefont
  {Pedersen}}, \bibinfo {author} {\bibfnamefont {D.~V.}\ \bibnamefont
  {Fedorov}}, \bibinfo {author} {\bibfnamefont {A.~S.}\ \bibnamefont
  {Jensen}},\ and\ \bibinfo {author} {\bibfnamefont {N.~T.}\ \bibnamefont
  {Zinner}},\ }\bibfield  {title} {\bibinfo {title} {Quantum few-body bound
  states of dipolar particles in a helical geometry},\ }\href
  {https://iopscience.iop.org/article/10.1088/0953-4075/49/2/024002} {\bibfield
   {journal} {\bibinfo  {journal} {J. Phys. B: At. Mol. Opt. Phys.}\ }\textbf
  {\bibinfo {volume} {49}},\ \bibinfo {pages} {024002} (\bibinfo {year}
  {2016})}\BibitemShut {NoStop}%
\bibitem [{\citenamefont {Siemens}\ and\ \citenamefont
  {Schmelcher}(2020)}]{siemensTunableOrderHelically2020}%
  \BibitemOpen
  \bibfield  {author} {\bibinfo {author} {\bibfnamefont {A.}~\bibnamefont
  {Siemens}}\ and\ \bibinfo {author} {\bibfnamefont {P.}~\bibnamefont
  {Schmelcher}},\ }\bibfield  {title} {\bibinfo {title} {Tunable order of
  helically confined charges},\ }\href
  {https://link.aps.org/doi/10.1103/PhysRevE.102.012147} {\bibfield  {journal}
  {\bibinfo  {journal} {Phys. Rev. E}\ }\textbf {\bibinfo {volume} {102}},\
  \bibinfo {pages} {012147} (\bibinfo {year} {2020})}\BibitemShut {NoStop}%
\bibitem [{\citenamefont {Plettenberg}\ \emph {et~al.}(2017)\citenamefont
  {Plettenberg}, \citenamefont {Stockhofe}, \citenamefont {Zampetaki},\ and\
  \citenamefont {Schmelcher}}]{plettenbergLocalEquilibriaState2017}%
  \BibitemOpen
  \bibfield  {author} {\bibinfo {author} {\bibfnamefont {J.}~\bibnamefont
  {Plettenberg}}, \bibinfo {author} {\bibfnamefont {J.}~\bibnamefont
  {Stockhofe}}, \bibinfo {author} {\bibfnamefont {A.~V.}\ \bibnamefont
  {Zampetaki}},\ and\ \bibinfo {author} {\bibfnamefont {P.}~\bibnamefont
  {Schmelcher}},\ }\bibfield  {title} {\bibinfo {title} {Local equilibria and
  state transfer of charged classical particles on a helix in an electric
  field},\ }\href {https://link.aps.org/doi/10.1103/PhysRevE.95.012213}
  {\bibfield  {journal} {\bibinfo  {journal} {Phys. Rev. E}\ }\textbf {\bibinfo
  {volume} {95}},\ \bibinfo {pages} {012213} (\bibinfo {year}
  {2017})}\BibitemShut {NoStop}%
\bibitem [{\citenamefont {Gloy}\ \emph {et~al.}(2022)\citenamefont {Gloy},
  \citenamefont {Siemens},\ and\ \citenamefont
  {Schmelcher}}]{gloyDrivenToroidalHelix2022}%
  \BibitemOpen
  \bibfield  {author} {\bibinfo {author} {\bibfnamefont {J.~F.}\ \bibnamefont
  {Gloy}}, \bibinfo {author} {\bibfnamefont {A.}~\bibnamefont {Siemens}},\ and\
  \bibinfo {author} {\bibfnamefont {P.}~\bibnamefont {Schmelcher}},\ }\bibfield
   {title} {\bibinfo {title} {Driven toroidal helix as a generalization of the
  {{Kapitza}} pendulum},\ }\href
  {https://link.aps.org/doi/10.1103/PhysRevE.105.054204} {\bibfield  {journal}
  {\bibinfo  {journal} {Phys. Rev. E}\ }\textbf {\bibinfo {volume} {105}},\
  \bibinfo {pages} {054204} (\bibinfo {year} {2022})}\BibitemShut {NoStop}%
\bibitem [{\citenamefont {Zampetaki}\ \emph {et~al.}(2013)\citenamefont
  {Zampetaki}, \citenamefont {Stockhofe}, \citenamefont {Kr{\"o}nke},\ and\
  \citenamefont {Schmelcher}}]{zampetakiClassicalScatteringCharged2013}%
  \BibitemOpen
  \bibfield  {author} {\bibinfo {author} {\bibfnamefont {A.~V.}\ \bibnamefont
  {Zampetaki}}, \bibinfo {author} {\bibfnamefont {J.}~\bibnamefont
  {Stockhofe}}, \bibinfo {author} {\bibfnamefont {S.}~\bibnamefont
  {Kr{\"o}nke}},\ and\ \bibinfo {author} {\bibfnamefont {P.}~\bibnamefont
  {Schmelcher}},\ }\bibfield  {title} {\bibinfo {title} {Classical scattering
  of charged particles confined on an inhomogeneous helix},\ }\href
  {https://link.aps.org/doi/10.1103/PhysRevE.88.043202} {\bibfield  {journal}
  {\bibinfo  {journal} {Phys. Rev. E}\ }\textbf {\bibinfo {volume} {88}},\
  \bibinfo {pages} {043202} (\bibinfo {year} {2013})}\BibitemShut {NoStop}%
\bibitem [{\citenamefont {Siemens}\ and\ \citenamefont
  {Schmelcher}(2025)}]{siemensClassicalScatteringFragmentation2025}%
  \BibitemOpen
  \bibfield  {author} {\bibinfo {author} {\bibfnamefont {A.}~\bibnamefont
  {Siemens}}\ and\ \bibinfo {author} {\bibfnamefont {P.}~\bibnamefont
  {Schmelcher}},\ }\bibfield  {title} {\bibinfo {title} {Classical scattering
  and fragmentation of clusters of ions in helical confinement},\ }\href
  {https://link.aps.org/doi/10.1103/PhysRevE.111.014140} {\bibfield  {journal}
  {\bibinfo  {journal} {Phys. Rev. E}\ }\textbf {\bibinfo {volume} {111}},\
  \bibinfo {pages} {014140} (\bibinfo {year} {2025})}\BibitemShut {NoStop}%
\bibitem [{\citenamefont {Zampetaki}\ \emph
  {et~al.}(2015{\natexlab{a}})\citenamefont {Zampetaki}, \citenamefont
  {Stockhofe},\ and\ \citenamefont
  {Schmelcher}}]{zampetakiDynamicsNonlinearExcitations2015}%
  \BibitemOpen
  \bibfield  {author} {\bibinfo {author} {\bibfnamefont {A.~V.}\ \bibnamefont
  {Zampetaki}}, \bibinfo {author} {\bibfnamefont {J.}~\bibnamefont
  {Stockhofe}},\ and\ \bibinfo {author} {\bibfnamefont {P.}~\bibnamefont
  {Schmelcher}},\ }\bibfield  {title} {\bibinfo {title} {Dynamics of nonlinear
  excitations of helically confined charges},\ }\href
  {https://link.aps.org/doi/10.1103/PhysRevE.92.042905} {\bibfield  {journal}
  {\bibinfo  {journal} {Phys. Rev. E}\ }\textbf {\bibinfo {volume} {92}},\
  \bibinfo {pages} {042905} (\bibinfo {year} {2015}{\natexlab{a}})}\BibitemShut
  {NoStop}%
\bibitem [{\citenamefont {Zampetaki}\ \emph
  {et~al.}(2015{\natexlab{b}})\citenamefont {Zampetaki}, \citenamefont
  {Stockhofe},\ and\ \citenamefont
  {Schmelcher}}]{zampetakiDegeneracyInversionBand2015}%
  \BibitemOpen
  \bibfield  {author} {\bibinfo {author} {\bibfnamefont {A.~V.}\ \bibnamefont
  {Zampetaki}}, \bibinfo {author} {\bibfnamefont {J.}~\bibnamefont
  {Stockhofe}},\ and\ \bibinfo {author} {\bibfnamefont {P.}~\bibnamefont
  {Schmelcher}},\ }\bibfield  {title} {\bibinfo {title} {Degeneracy and
  inversion of band structure for {{Wigner}} crystals on a closed helix},\
  }\href {https://link.aps.org/doi/10.1103/PhysRevA.91.023409} {\bibfield
  {journal} {\bibinfo  {journal} {Phys. Rev. A}\ }\textbf {\bibinfo {volume}
  {91}},\ \bibinfo {pages} {023409} (\bibinfo {year}
  {2015}{\natexlab{b}})}\BibitemShut {NoStop}%
\bibitem [{\citenamefont {Liu}\ and\ \citenamefont
  {Nocedal}(1989)}]{liuLimitedMemoryBFGS1989}%
  \BibitemOpen
  \bibfield  {author} {\bibinfo {author} {\bibfnamefont {D.~C.}\ \bibnamefont
  {Liu}}\ and\ \bibinfo {author} {\bibfnamefont {J.}~\bibnamefont {Nocedal}},\
  }\bibfield  {title} {\bibinfo {title} {On the limited memory {{BFGS}} method
  for large scale optimization},\ }\href {https://doi.org/10.1007/BF01589116}
  {\bibfield  {journal} {\bibinfo  {journal} {Math. Program.}\ }\textbf
  {\bibinfo {volume} {45}},\ \bibinfo {pages} {503} (\bibinfo {year}
  {1989})}\BibitemShut {NoStop}%
\bibitem [{\citenamefont {Byrd}\ \emph {et~al.}(1995)\citenamefont {Byrd},
  \citenamefont {Lu}, \citenamefont {Nocedal},\ and\ \citenamefont
  {Zhu}}]{byrdLimitedMemoryAlgorithm1995}%
  \BibitemOpen
  \bibfield  {author} {\bibinfo {author} {\bibfnamefont {R.~H.}\ \bibnamefont
  {Byrd}}, \bibinfo {author} {\bibfnamefont {P.}~\bibnamefont {Lu}}, \bibinfo
  {author} {\bibfnamefont {J.}~\bibnamefont {Nocedal}},\ and\ \bibinfo {author}
  {\bibfnamefont {C.}~\bibnamefont {Zhu}},\ }\bibfield  {title} {\bibinfo
  {title} {A {{Limited Memory Algorithm}} for {{Bound Constrained
  Optimization}}},\ }\href {https://epubs.siam.org/doi/10.1137/0916069}
  {\bibfield  {journal} {\bibinfo  {journal} {SIAM J. Sci. Comput.}\ }\textbf
  {\bibinfo {volume} {16}},\ \bibinfo {pages} {1190} (\bibinfo {year}
  {1995})}\BibitemShut {NoStop}%
\end{thebibliography}
\end{document}